\documentclass[fleqn,10pt]{wlscirep}
\usepackage{amsfonts, amssymb, amsmath, amsthm, mathrsfs}
\usepackage[utf8]{inputenc}
\usepackage{verbatim}
\newtheorem{theorem}{Theorem}
\newtheorem{lemma}[theorem]{Lemma}

\newcommand{\ket}[1]{|#1 \rangle}
\newcommand{\bsigma}{\boldsymbol{\sigma}}
\usepackage{algorithm,algpseudocode}

\usepackage{algorithm}
\usepackage{algpseudocode}
\usepackage{setspace}
\newcommand\blfootnote[1]{%
  \begingroup
  \renewcommand\thefootnote{}\footnote{#1}%
  \addtocounter{footnote}{-1}%
  \endgroup
}
\title{Quantum State Preparation of Normal Distributions using Matrix Product States}
\author[1]{Jason Iaconis}
\author[1]{Sonika Johri}
\author[2]{Elton Yechao Zhu}
\affil[1]{IonQ Inc, 4505 Campus Dr, College Park, MD 20740, USA}
\affil[2]{Fidelity Center for Applied Technology,
FMR LLC, Boston, MA 02210, USA}
\begin{abstract}
State preparation is a necessary component of many quantum algorithms. In this work, we combine a method for efficiently representing smooth differentiable probability distributions using matrix product states with recently discovered techniques for initializing quantum states to approximate matrix product states. Using this, we generate quantum states encoding a class of normal probability distributions in a trapped ion quantum computer for up to 20 qubits. We provide an in depth analysis of the different sources of error which contribute to the overall fidelity of this state preparation procedure. Our work provides a study in quantum hardware for scalable distribution loading, which is the basis of a wide range of algorithms that provide quantum advantage.
\end{abstract}
\begin{document}
\flushbottom
\maketitle

\section{Introduction}
\blfootnote{Fidelity Public Information}
Quantum state preparation is an important step in many quantum
algorithms such as Monte-Carlo methods for quantum computers \cite{Montanaro2015},
quantum algorithms for systems of linear equations \cite{Harrow2009}, various
quantum machine learning algorithms\cite{Kerenidis2017,Mitarai2018} and
Hamiltonian simulation \cite{Childs2018,Low2019hamiltonian} which are expected
to have broad applications in a variety of fields.  In order for these algorithms to provide quantum
advantage, this state preparation procedure must be performed efficiently and
with sufficiently low noise. 

Previously, in the literature, many near-term techniques have been proposed to realize quantum state preparation, including numerical integration \cite{Grover2022, Carrera2021} and quantum generative adversarial networks \cite{Zoufal2019,Zhu2022}. Refs. \citeonline{Carrera2021,Zoufal2019,Zhu2022} also experimentally implemented their state preparation protocols. However, there is a lack of study into their efficiency. There have also been fault-tolerant proposals for quantum state preparation \cite{Sun2021,Zhang2022,Rattew2022,McArdle22}, with theoretical guarantees in terms of convergence, circuit depth, and the number of ancillas. However, due to their high resource requirements, these techniques cannot be realized experimentally on near-term devices. On the other hand, Refs. \citeonline{Sanders19,Bausch2022} considered black-box state preparation which assumes the existence of an oracle that produces target coefficients. This unfortunately does not allow an end-to-end implementation.

In this work, we make use of matrix product states (MPS) which are an interesting class of quantum wave functions whose probability amplitudes can be described using a specific tensor structure. MPS based algorithms are some of the most powerful classical numerical techniques for manipulating large scale quantum states \cite{Schollwock_2011}. These states were originally studied in the context of the density matrix renormalization group (DMRG) algorithm \cite{white1992density,PhysRevB.48.10345} and used mainly for simulating the ground states of local many-body quantum systems. The key property is that the amount of entanglement is the relevant quantity which controls the efficiency of the MPS representation \cite{VidalLatoreKitaev, latorre2004ground}. Since their initial discovery, these tensor network structures and their associated algorithms have grown in popularity and have been applied to many different areas of quantum information theory \cite{Vidal2003, Verstraete2006}.

In Ref. \citeonline{garcia2021quantum}, it was realized that  when the amplitudes of a quantum state can be described by a smooth function, the entanglement of the state grows very slowly with increasing qubit number. These properties of MPS states and smooth wave functions were used in Ref. \citeonline{Holmes2020}, to describe an efficient method of approximating probability distributions using matrix product states and quantum circuits with a single layer of two-qubit entangling unitaries. Recently, efforts have been made to prepare matrix product state wave functions to greater accuracy with more general short depth quantum circuits. In particular, Ref. \citeonline{ran2020encoding} developed an iterative algorithm which can approximate high bond dimension MPS states using short depth circuits composed of  two-qubit unitary gates.

In this paper, we propose and experimentally demonstrate preparation of quantum states which encode normal distributions as amplitudes. Importantly, we do a detailed analysis of the theoretical and experimental error of our protocol.  Our general procedure involves several steps. First we approximate the normal distribution with a low-degree piece-wise polynomial function given by the Irwin-Hall distributions. We then use the procedure given in Ref. \citeonline{Oseledets2013} to exactly prepare a low bond dimension matrix product state which encodes this piece-wise polynomial distribution. Next, use the iterative MPS loading procedure described in Ref. \citeonline{ran2020encoding} to create a controlled low-depth quantum circuit which prepares an approximation to the MPS circuit. Finally, we execute the circuit and measure the resulting probability distribution.

 The rest of this paper is organized as follows. In section \ref{subsec:statepreparation}-\ref{subsec:distance}, we review the matrix product state formalism and the iterative method which allows us to prepare MPS states with low depth quantum circuits. In section \ref{subsec:IrwinHallError}-\ref{sec:resource} we provide a rigorous theoretical analysis of the sources of error contributing to the difference between the ideal probability distribution and the one produced by our procedure. In section \ref{subsec:MC}, we analyze the applicability to Monte Carlo integration and discuss the limitations in asymptotic scaling in terms of the error tolerance. In section \ref{subsec:exp}, we show the result of applying our state preparation procedure to prepare quantum wave functions on up to 20 qubits using a state-of-the-art  trapped ion quantum computer. Finally, we conclude by summarizing our results and discussing potential applications of this work in section \ref{sec:discussion}.

\section{Results}\label{sec:MPS}

\subsection{Quantum State Preparation}\label{subsec:statepreparation}

The main objective of the state preparation procedure is to generate a quantum circuit which initializes a wavefunction on $N$ qubits,
\begin{eqnarray}
    \ket{\Psi} = \sum_{x=0}^{2^N-1} c_x \ket{x},
\end{eqnarray}
whose amplitudes $c_x$ are set to a specified value.  It is known that in general, such a circuit must include a number of gates that scales exponentially with $N$ \cite{Knill1995}. This is true even if the goal is to only approximately prepare a state $\ket{\Psi'}$ to some fixed precision $\epsilon$ such that $||\ket{\Psi'}-\ket{\Psi}|| \leq \epsilon$, where $\|\cdot\|$ is understood as the L2 norm.  

However, if we restrict ourselves to certain families of wave functions with specific structure, it becomes possible to efficiently prepare the state with a circuit which contains only a number of quantum gates that scales polynomially with $N$. For instance, Grover and Rudolf \cite{Grover2022} gave a procedure for efficiently preparing wavefunctions whose amplitudes are described by an efficiently integrable probability distribution. However, Ref. \citeonline{Herbert2021} pointed out that the Grover-Rudolf method does not give the quadratic speedup in quantum Monte-Carlo algorithm when classical Monte-Carlo integration is used to determine the gate angles in the state preparation circuit. Ref. \citeonline{Rattew2021efficient} presented a quantum algorithm to efficiently prepare normal distributions using Mid-Circuit Measurement and Reuse. Ref. \citeonline{Zhang2022} presented a quantum algorithm that prepares any $N$-qubit quantum state with $\Theta(N)$-depth circuit, with an exponential amount of ancillary qubits. Ref. \citeonline{McArdle22} described a state preparation algorithm that uses quantum eigenvalue transformation, and obtains promising gate complexity. However, the actual Toffoli gates count of $O(10^4)$ means it will remain unreachable in the near future.

In Ref. \citeonline{Holmes2020}, it was shown that if the probability distribution to be encoded is a smooth differentiable function, then there exists an efficient preparation scheme based on an encoding of the probability distribution as a matrix product state tensor network. Matrix product state wave functions are a class of quantum states on which efficient classical computations can be performed even when the number of qubits in the system is large. As we will describe in this section, there exist efficient methods of generating MPS representations of low degree polynomial functions, as well as methods for approximately preparing MPS wave functions using low depth quantum circuits. In this work, we use this encoding scheme to prepare states which approximate normal probability distributions. With this technique, we are also able to make use of the well known family of piece-wise polynomial functions known as the Irwin-Hall distributions which approximate the normal distribution to arbitrary accuracy.

\subsection{MPS formalism} \label{subsec:MPS_formalism}

A Matrix Product State (MPS) is a wave function of the form

\begin{eqnarray}
    \ket{\Psi} =
    \sum_{\{\sigma\}}M^{[1],\sigma_1}_{\alpha_1}M^{[2],\sigma_2}_{\alpha_1,\alpha_2}
    \dots  M^{[N-1],\sigma_{N-1}}_{\alpha_{N-2},\alpha_{N-1}}
    M^{[N],\sigma_N}_{\alpha_{N-1}} \ket{\sigma_1 \sigma_2 \dots
    \sigma_{N-1}\sigma_N}, \label{eq:mps}
\end{eqnarray}
where the terms $M^{[i],\sigma_i}_{\alpha_{i-1},\alpha_i}$ are $N$ different 3-index tensors, and we use the Einstein summation convention that repeated indices are summed over. Each tensor contains a `physical' index $\sigma_i \in [1,d]$, and `bond' indices $\alpha_i \in [1,\chi]$ \cite{Schollwock_2011}. Here $d$ is the local dimension of the quantum state, so that $d=2$ for qubits.

We call the maximum value of the bond index, $\chi$, the bond dimension of the MPS.
Wave functions which can be represented by a bond-dimension $\chi$
MPS can be completely defined using only $d*N$ matrices of dimension $\chi\times\chi$ and
therefore can be stored using only $d*N*\chi^2$ complex numbers instead of storing all $2^N$ complex amplitudes directly.

Only a small subset of wave functions can be represented exactly with a finite bond-dimension MPS.  In particular, MPS wave functions are very good approximations for quantum states with low entanglement. The entanglement entropy, $S=-Tr[\rho_A\log\rho_A]$, of a MPS wave function with fixed bond dimension $\chi$ is bounded by $S\le \log(\chi)$.   

In general, if we let $\chi = 2^N$, we can represent any wavefunction $\ket{\psi}$ using the MPS form given by Eq~\ref{eq:mps}. One of the most important features of matrix product states is the ability to compress such a wavefunction in a controlled manner to create a fixed small $\chi$ approximation  to $\ket{\psi}$. The most straightforward approach to this is the singular value decomposition (SVD) compression scheme, described in Ref. \citeonline{Schollwock_2011}. We start with the wave function whose coefficients are given by the $N$ component tensor $\psi_{\sigma_1,\sigma_2,\dots,\sigma_N}$ such that
\begin{eqnarray}
\ket{\psi} = \sum_{\sigma_1,\sigma_2,\dots \sigma_N} \psi_{\sigma_1,\sigma_2,\dots\sigma_N}\ket{\sigma_1,\sigma_2,\dots\sigma_N}.
\end{eqnarray}
This tensor is reshaped into a rectangular matrix $\psi_{\sigma_1,(\sigma_2 \dots \sigma_N)}$.  A singular value decomposition on this matrix allows us to write

\begin{align}
\psi_{\sigma_1,(\sigma_2 \dots \sigma_N)} & = \sum_{a_1=1}^m U_{\sigma_1,a_1}S_{a_1,a_1} (V^\dagger)_{a_1,(\sigma_2...\sigma_N)} \\
& = \sum_{a_1=1}^m U_{\sigma_1,a_1} \psi_{a_1, \sigma_2,\sigma_3,\dots,\sigma_N}.\nonumber
\end{align}
We restrict the sum to include only the largest $m$ singular values of the diagonal matrix $S$.  We can now reshape the remaining tensor to form the matrix $\psi_{(a_1 \sigma_2),(\sigma_3\dots \sigma_N)}$. A singular value decomposition is applied to this matrix and the process is repeated for all qubits $i$, resulting in the decomposition given above, where each matrix $U = M_{\alpha_i,\alpha_{i+1}}^{[i],\sigma_i}$. For matrix dimension $(m\times n)$ with $m>n$, the cost of the SVD is $\Tilde{O}(mn^2)$, which implies that the SVD costs $\mathcal{O}(N\chi^2\chi')$ when truncating from bond dimension $\chi$ to bond dimension $\chi'$. There also exist more complex compression techniques \cite{Schollwock_2011} which may be more effective in certain cases. For example, the iterative variational compression algorithm which solves a series of $\chi'^2\times \chi'^2$ linear equations which depends only on the truncated bond-dimension $\mathcal{O}(\chi')$. Although deriving these equation involves contracting over the original bonds of size $\chi$, which has cost $\mathcal{O}(\chi'\chi^2)$, this method can lead to a large practical speedup in certain situations..

For each matrix, the so called truncation error is given by the sum of the squares of the discarded singular values $\epsilon^2 = \sum_{i=m+1}^{2^N} \lambda_i^2$, and controls the fidelity of this compression method \cite{Schollwock_2011}. For a state $\ket{\psi}$ and allowed error $\epsilon$, we say that it can be approximately represented as a MPS if, for arbitrary $N$, there exists a fixed $\chi$ MPS,
$\ket{\tilde{\psi}}$ such that
\begin{eqnarray}\label{eq:epsilon}
    ||\ket{\psi} - \ket{\tilde{\psi}}|| \le \epsilon.
\end{eqnarray}
In this case, efficient classical computations can be performed on $\ket{\tilde{\psi}}$.

\subsection{Smooth Differentiable Functions}
It turns out that many states that are input to quantum algorithms inherently possess a low degree of entanglement and therefore can be represented using matrix product states. For example, common distributions used in Monte Carlo methods include the uniform distribution, the normal distribution, and the log-normal distribution.

Most importantly for our purposes, smooth differentiable real-valued functions which are appropriately encoded into the amplitude of quantum states satisfy this low entanglement property. Consider a normalized real smooth probability distribution $f(x)$ defined on an interval $[a,b]$. A discretized version of this function can be encoded into the amplitudes of an $N$-qubit quantum register. Specifically, throughout this work, we use the big-endian binary encoding on the interval $[a,b]$, such that
\begin{eqnarray}
x_k = a + \frac{b-a}{2^{N}-1}k \, = \, a + \frac{k}{h}
\label{eq:discrete_var}
\end{eqnarray}

where $h=(2^N-1)/(b-a)$ and $k = k(\bsigma)$ can be represented by a binary bit-string $\bsigma = \sigma_1\sigma_2 \dots \sigma_N$ so that 
\begin{eqnarray}
    k = \sum_i \sigma_i 2^{N-i}.
\end{eqnarray}

Therefore, the discretized amplitude encoded wave function takes the form
\begin{eqnarray}
\ket{\psi} = \sum_{k=0}^{2^{N}-1} \sqrt{f(x_k)}\ket{k},  \label{eq:amp_wf}
\end{eqnarray}
where $\ket{k}$ is the integer representation of the computational basis state $\ket{\bsigma}$ in big-endian notation. For a fixed number of qubits $N$, the wave function $\ket{\psi}$ encodes a discretized version of the probability distribution $f(x)$, sampled at the discrete points $x_k$.

Consider the effect of adding one additional qubit to the state $\ket{\psi}$ so that $\ket{\sigma_1 \sigma_2 \dots \sigma_{N}} \rightarrow \ket{\sigma_1\sigma_2\dots \sigma_{N} \sigma_{{N}+1}}$.  Within the big-endian binary encoding scheme, adding one additional quantum register doubles the density of the discretized points in $f(x_k)$.  Furthermore, since the added qubit encodes the least significant bit of $k$, the location of the previous $2^{N}$ points is not changed by adding one more qubit to the system.

In Ref.~\citeonline{garcia2021quantum}, the entanglement entropy of these discretized amplitude encoded wave functions was carefully analyzed. It was found that adding one additional qubit leads to only a small increase in the entanglement of the state $\ket{\psi}$, and that this change is controlled by the maximum value of the derivative of $f(x)$
\begin{eqnarray}
\Delta_D =\max_{x}|f^\prime(x)|.
\end{eqnarray}
Specifically, adding one additional qubit increases the entanglement entropy $S(\rho)$ of the final state as
\begin{eqnarray}
\Delta S[\rho^{({N})}] \le \mathcal{O}\left (2\sqrt{\Delta_D}|b-a|2^{-{N}/2}\right).
\end{eqnarray}
Therefore, for each additional qubit we add to our encoding, the discretization error is cut in half, but only a vanishingly small amount of entanglement is added to the system.

As shown in Ref.~\citeonline{Holmes2020}, the slow scaling of the entanglement entropy with system size open up the possibility of efficiently representing wave functions of the form in Eq.~\ref{eq:amp_wf} using Matrix Product States. For a given probability distribution $f(x)$, we can create a MPS representation of $\ket{\psi}$ by setting the elements of $M_{\alpha_i,\alpha_j}^{\sigma_j}$ in Eq.~\ref{eq:mps} to appropriate values. The elements of $M$ can be exactly determined by performing the repeated singular-value decomposition on the wave function amplitudes $\psi_{\sigma_1,\sigma_2,\dots \sigma_{N}} = f(x_{(k(\bsigma))})$. However, for large $N$, storing the tensor for original $\psi$ becomes computationally intractable.

Instead, we can efficiently determine the elements of $M$ for arbitrarily large ${N}$ if our function $f(x_k)$ is a degree-$p$ polynomial function $f(x)= \sum_{i=0}^p a_i x_i $. In this case, Ref.~\citeonline{Oseledets2013} explicitly gives the elements of the matrices $M$ in terms of the coefficients $a_i$. 
For this, first we define the auxiliary variables $t_i(\sigma_i)$, defined for a given bit-string $k=k(\bsigma)$. 
\begin{eqnarray}
t_i = t(\sigma_i) &=& a \delta_{i,1} + \frac{2^{i-1}}{h} \sigma_i \\
x &=& t_1 + \dots + t_{N}.
\end{eqnarray}

Now we can write the elements of $M_{\alpha_i,\alpha_j}^{[j]\sigma_j}$ as
\begin{eqnarray}
\phi_s(x) &=& \sum_{k=s}^p a_k C_k^s x^{k-s} \label{eq:mps_fcn_enc_1}
\\
M^{[1]\sigma_1}_{\alpha_1} &=&  \phi_{\alpha_1}(t_1) \\
M^{[j]\sigma_j}_{\alpha_i,\alpha_j} &=&
\left \{ \begin{array}{cc}
C_{\alpha_i}^{\alpha_i-\alpha_j} t_{\alpha_j}^{\alpha_i-\alpha_j} &\, \text{ if } \, \alpha_i>\alpha_j\\
0  &\, \text{ if } \, \alpha_i<\alpha_j
\end{array} \right . \\
M_{\alpha_{N}}^{[N]\sigma_{N}} &=& t_N^{\alpha_N}
\label{eq:mps_fcn_enc_4}
\end{eqnarray}
where $C_i^j$ is the binomial coefficient and where all bond indices $\alpha_i \in (0,\dots, p)$. These equations are directly adapted from Theorem 6 of Ref.~\citeonline{Oseledets2013}.  Notice that for a degree-$p$ polynomial, we must calculate only $2\times {N}\times p^2$ coefficients, where half of these coefficients are zero, giving us an efficient representation of the function $f(x)$ in the MPS representation. 

Furthermore, in Ref.~\citeonline{Holmes2020}, a method was presented for extending this encoding to piece-wise degree-$p$ polynomial functions when the domain is a fraction of the full domain $[a,b]$. In this case, we wish to encode the function into $2^k$ separate regions, where the function of the $\ell^{\text{th}}$ region is given by
\begin{eqnarray}
f_\ell(x)  &=& \sum_{k=0}^p a_{k}^{(\ell)} x^k \hspace{5mm} \text{ for } \quad a+\ell \frac{2^k}{h} < x < a + (\ell+1)\frac{2^k}{h} \\
&=&  0 \hspace{18mm} \text{ otherwise} . 
\end{eqnarray}
 
To encode this function, we first represent $f_\ell(x)$ as an MPS on the full domain $[a,b]$, using Eq.'s~\ref{eq:mps_fcn_enc_1}-\ref{eq:mps_fcn_enc_4}, and then `zero out' elements of the tensors $M_{\alpha_i \alpha_j}^{[j]\sigma_j}$  corresponding to regions outside the domain of region $\ell$. To do this, first let $\ell$ be represented by a binary bit-string $\ell = b_1 b_2\dots b_k$. Then, we set
\begin{eqnarray}
M_{\alpha_i,\alpha_j}^{[j],1} = 0 \text{ if } b_j=0 \\
M_{\alpha_i,\alpha_j}^{[j],0} = 0 \text{ if } b_j=1 .
\end{eqnarray}
All matrices $M_{\alpha_i,\alpha_j}^{[j],\sigma_j}$ for $j>k$ are left unchanged. In other words, if we replace all the  matrices in Eq.~\ref{eq:mps}, $M_{\alpha_i,\alpha_j}^{[j],\sigma_j}$, with a zero matrix of equal size if $b_j\neq \sigma_j$, it will have the effect of setting all amplitudes outside the domain of region $\ell$ to zero. 

Finally, we can put all these pieces together to write an efficient encoding scheme for representing a piece-wise degree-$p$ polynomial function of $2^k$ separate domains as a low bond dimension MPS state. For each sub-region on the full domain $[a,b]$, we encode $f_\ell(x)$ into the MPS ${\bf M}_\ell$. We then use the property that two Matrix Product States, $M_1$ and $M_2$ with bond dimensions $\chi$ and $\chi^\prime$ can be added together to form a MPS $M_3 = M_1+M_2$ of bond dimension $\chi+\chi^\prime$.  Therefore, we may add all Matrix Product States on the $2^k$ sub-domains together to get a final representation ${\bf M}_T = \sum_{\ell=0}^{2^k} M_\ell$, which is a MPS of bond dimension $2^k (p+1)$. ${\bf M}_T$ is an efficient approximation to the target quantum state which we wish to prepare using a quantum device. The procedure for generating   ${\bf M}_T$ is summarized in Algorithm \ref{alg:mps_enc}.

\begin{algorithm}
\caption{MPS Encoding Procedure}
\hspace*{\algorithmicindent} 
\begin{algorithmic}[1]
\Require{ A degree-$p$ piece-wise function  $f_\ell(x)  = \sum_{j=0}^p a_{j}^{(\ell)} x^j $. System size ${N}$. Domain [a,b]. Support bit $k$.} 
\Ensure{A $\chi \le 2^k(p+1)$ MPS, $\bf{M}_T$ which encodes $f_\ell(x)$}
\Statex
\For{$\ell \gets 1$ to $2^k$}   
    \State {Encode $f_\ell(x)$ into ${\bf M}_\ell$ on domain [a,b]}
    \State {Zero out  ${\bf M}_\ell$ outside domain $D_\ell$}
    \EndFor
\State \Return{ ${\bf M}_T \leftarrow  \sum_{\ell=0}^{2^k} {\bf M}_\ell$}

\end{algorithmic}

\label{alg:mps_enc}
\end{algorithm}

\subsection{Quantum Circuits for MPS States}

We now describe how we can generate low depth quantum circuits which prepare MPS wave functions. A bond dimension $\chi$ MPS state can be exactly created with a quantum circuit which is composed of $N$ local unitary operators each acting on $m=\log(\chi)+1$ qubits. While in principle this allows us to efficiently map a given MPS to a polynomial depth quantum circuit, the constant factor overhead of compiling arbitrary $m$-qubit quantum gates to a basic set of one and two qubit gates quickly becomes infeasible for NISQ devices.

\begin{figure}[t]
    \centering
    {\bf a)} \,\, \includegraphics[height=1.3in]{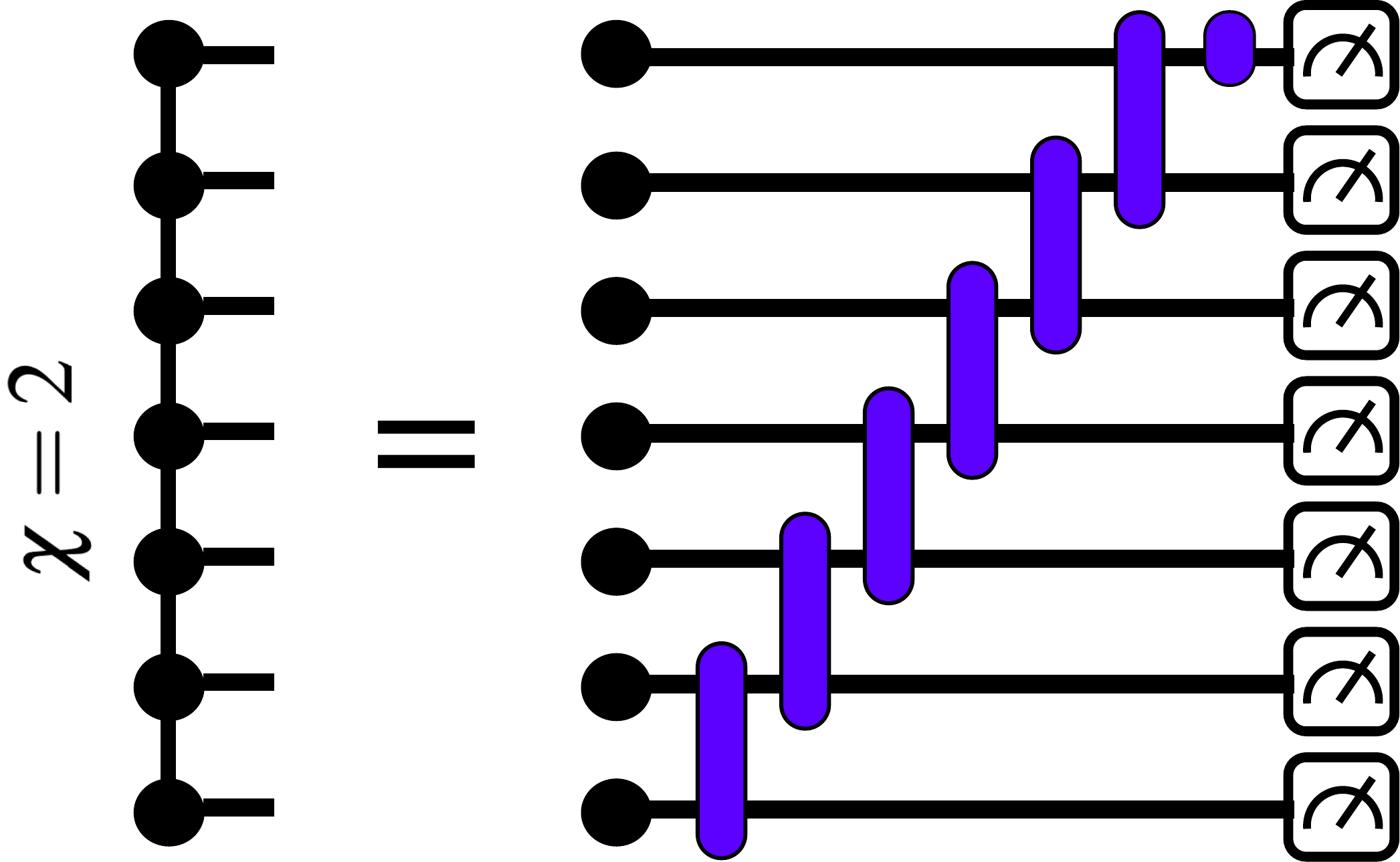}
 \hspace{25mm}     {\bf b)} \, \includegraphics[height=1.3in]{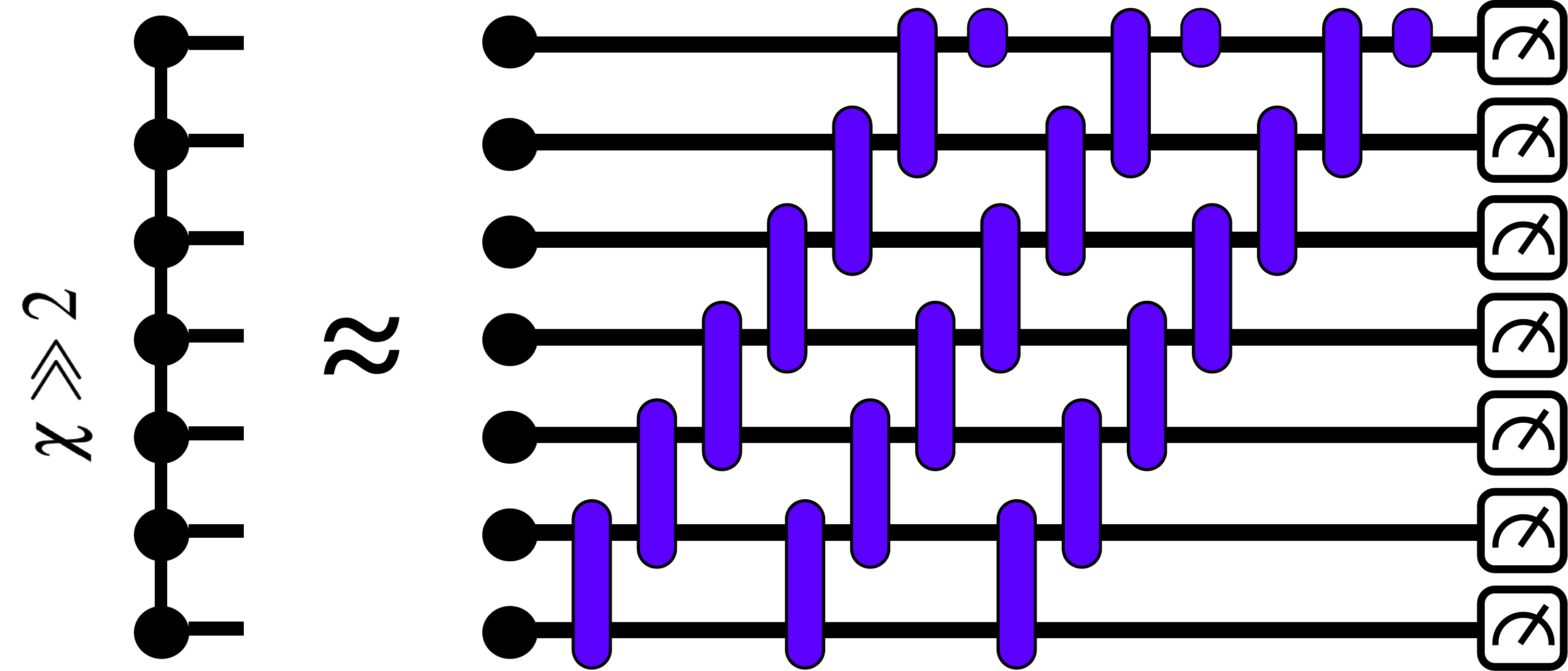} \\
\vspace{3pc}
    {\bf c)}\,\, \includegraphics[width=2.8in]{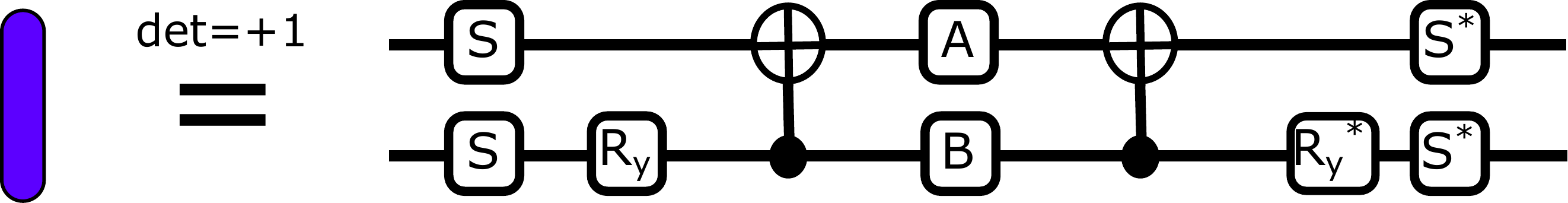}
     \hspace{16mm} 
    {\bf d)} \includegraphics[width=2.8in]{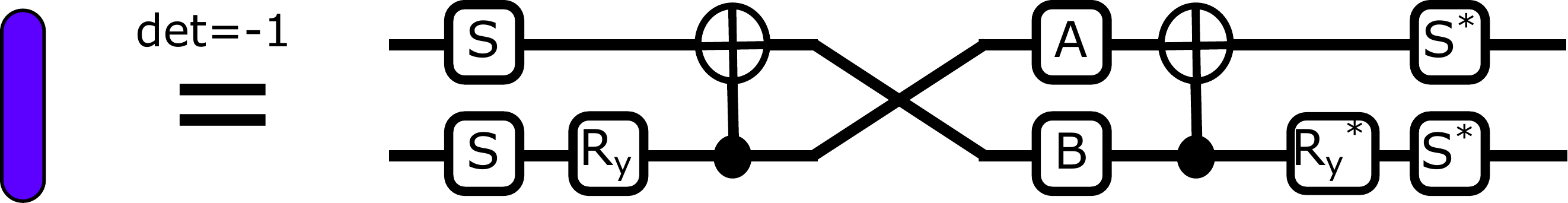}
    \caption{{\bf The circuit structure for preparing an MPS wave function.} {\bf a)} A MPS state with all bond dimensions equal to 2 can exactly be contructed using (N-1) 2-qubit unitary operators plus one single qubit rotation, arranged in the pictured order. {\bf b)} A higher bond dimension MPS state can be approximately prepared by repeatedly applying single layer MPS approximations, following the construction of Ref.~\citeonline{ran2020encoding}. {\bf c)} If the entries of the MPS state are real, then all 2-qubit gates $U$ are O(4) operators and can be implemented using single qubit rotations and only 2 CNOT gates if $\det(U)=+1$, or {\bf d)} 2 CNOT gates plus a SWAP operator if $\det(U)=-1$.}
    \label{fig:mps_circuit}
\end{figure}

Consequently, a number of proposals for approximately generating bond dimension=$\chi$ MPS states using low depth quantum circuits have been put forward \cite{ran2020encoding, zapata_mps, shirakawa, PhysRevX.12.011047, PRXQuantum.2.010342}.  In this work, we take the iterative approach of Ref. \citeonline{ran2020encoding}. In this method, a high bond dimension MPS wave function is iteratively approximated by applying $D$ layers of local unitary operators.  Each layer of gates is designed to prepare a $\chi=2$ MPS wave function and can therefore be efficiently prepared using only two-qubit unitary operators.  The  procedure for constructing the specific set two-qubit unitary operators which exactly prepares a $\chi=2$ MPS state is described in Ref. \citeonline{ran2020encoding}, and results in a circuit architecture of the form shown in Fig.~\ref{fig:mps_circuit} a).  When several of these layers are combined together, a circuit in the form of Fig.~\ref{fig:mps_circuit} b) is able to approximately prepare a $\chi\gg 2$ MPS wave function.  The key idea of the algorithm is that each layer of gates also acts as a ``disentangler'' operator which can be applied to the original wavefunction.  The algorithm proceeds as follows

\vspace{1pc}

\begin{algorithm}[H] 
\caption{Iterative Circuit Preparation}
\begin{algorithmic}[1]
\Require{ A target MPS function $\ket{\psi_0}$} 
\Ensure{A quantum circuit $U_{tot} = U_0 U_1\dots U_{D-1} $}

\Statex
\For{$i \gets 0$ to $D-1$} 

    \State {Truncate $\ket{\psi_{i}}$ to form a bond dimension 2 MPS $\ket{\tilde{\psi}_i}$}
    \State {Generate $U_i$ s.t. $U_i \ket{0} = \ket{\tilde{\psi}_i}$ }
    \State {Generate $\ket{\psi_{i+1}} = U_i^\dagger \ket{\psi_{i}}$}
\EndFor

\State \Return {$U_{tot} = U_0U_1\dots U_{D-1}$}
\end{algorithmic}
\end{algorithm}

The truncation procedure applied in Algorithm 2 is the SVD compression method described in section \ref{subsec:MPS_formalism}. Each layer of unitary gates, $U_i$ approximately prepares the target wave function $\ket{\psi_{i-1}}$.  Therefore, the adjoint circuit layer $U_i^\dagger$ will approximately take the target $\ket{\psi_{i-1}}$ to the product state $\ket{0}$. After each iteration, the entanglement of the wave function $\ket{\psi_{i}}$ is therefore lower than the previous wave function $\ket{\psi_{i-1}}$. In this way, the bond dimension-2 approximations $\ket{\tilde{\psi}}$ become increasingly accurate.  The final state we produce is
\begin{eqnarray}
\ket{\tilde{\psi}} &=& U_0 U_1 \dots U_{D-1}\ket{0}
\end{eqnarray}

The fidelity of this state preparation procedure is given by the expression.
\begin{align}
    \langle \psi_0 \ket{\tilde{\psi}} &= \langle \psi_0 |U_0 U_1 \dots U_{D-1} \ket{0} \\
    &= \langle{\psi_1} | U_1 \dots U_{D-1} \ket{0} = \langle{\psi_{D-1}}|U_{D-1} |0\rangle .\nonumber
\end{align}
Therefore, the accuracy of this preparation method depends on the ability of the unitary operators to disentangle the wave function. In section \ref{sec:theory}, we look at how the approximation error decreases as the number of layers increases for wave functions described by a normal distribution.  For these wave functions, a small number of layers is generally sufficient to prepare a good approximate wave function. Therefore, using the techniques described in this section, we can efficiently create low depth quantum circuits which approximately prepare the target normal distribution wave functions on a large number of qubits.  After executing this circuit, a projective measurement on all qubits generates a single bit-string, which can converted, using Eq.~\ref{eq:discrete_var}, to one of $2^N$ discrete values on the interval $[a,b]$. Therefore, sampling many output bit-strings generated by the quantum circuit allows one to sample real numbers from the probability distribution encoded in the quantum wave function.

\subsection{Irwin-Hall Distribution}
The Irwin-Hall distribution is the continuous probability distribution for the sum of $n$ i.i.d. $\mathcal{U}(0,1)$ random variables,
\begin{equation}
    X_n=\sum_{k=1}^n \mathcal{U}_k.
\end{equation}
Here $\mathcal{U}(0,1)$ is the uniform random variable over $[0,1]$.

The probability density function (pdf) is given by
\begin{equation}\label{eq:irwinhallpdf}
    f_{X_n}(x)=\frac{1}{2(n-1)!}\sum_{k=0}^n(-1)^k\binom{n}{k}(x-k)^{n-1}\text{sgn}(x-k),
\end{equation}
where $\text{sgn}(x-k)$ denotes the sign function
\begin{equation*}
    \text{sgn}(x-k)=
    \begin{cases}
    -1\hspace{2em} &x<k\\
    0\hspace{2em} &x=k\\
    1\hspace{2em} &x>k.
    \end{cases}
\end{equation*}
From here, we see that the pdf of $X_n$ is piece-wise polynomial, with $n$ pieces, and each polynomial is of degree $n-1$.

By the Central Limit Theorem, as $n$ increases, the Irwin-Hall distribution converges to a normal distribution with mean $\mu=n/2$ and variance $\sigma^2=n/12$. Formally, it means
\begin{equation}
    \sqrt{12/n}(X_n-n/2)\to \mathcal{N}(0,1) \hspace{2em}\text{in distribution},
\end{equation}
where $\mathcal{N}(\mu,\sigma^2)$ denotes the normal distribution with mean $\mu$ and variance $\sigma^2$. For a sequence of real-valued random variables, convergence in distribution means
\begin{equation}
    \lim_{n\to \infty}F_n(x)=F(x)\hspace{2em} \forall x\in R
\end{equation}
at which $F$ is continuous. Here $F_n,F$ denote cumulative distribution functions (cdf).
In Section \ref{subsec:IrwinHallError}, we present a more rigorous convergence result in pdf.
\subsection{Distance Measures and Statistical Analysis}\label{subsec:distance}
We use common distance measures such as the L1 and sup norm between the pdf/cdf of the Irwin-Hall distributions and the corresponding normal distributions. This can be done when the pdf/cdf is available analytically or numerically. When experiments are performed, there is no access to the intrinsic quantum state, as only measurement samples are collected. In this case, we use the 1-sided Kolmogorov-Smirnov test, and obtain the KS statistics, defined as
\begin{equation}
    D_n=\sup_x|F_n(x)-\tilde{F}(x)|.
\end{equation}
Here $F_n(x)$ is the empirical cdf derived from the samples, and $\tilde{F}$ is the cdf of the ideal distribution to which it is compared. The KS statistics can be thought of as an approximation of the sup norm between the cdfs.  This KS statistic can also be used to test whether the two underlying distributions $F_n(x)$ and $\tilde{F}(x)$ differ from each other with statistical significance. When a finite number of samples $s$ are taken from both distibutions, the null hypothesis that $F_n(x)=\tilde{F}(x)$ is accepted if 
\begin{equation}
    D_n < \sqrt{\ln\left(\frac{2}{\alpha}\right)\frac{1}{s}},
    \label{eq:KS_test}
\end{equation}
where $\alpha$ is the significance level which is commonly taken to be $\alpha=0.05$. 

We note that there are alternative statistical test methods such as the Shapiro-Wilk test or the Anderson-Darling test. However, their test statistics are not as useful as the KS statistics.

A thorough understanding of the error analysis associated with state preparation is important. In an application such as Monte Carlo methods, one often needs to obtain results to a desired accuracy. Therefore the state preparation needs to prepare the corresponding distribution to a precision that is compatible with the desired application.
\label{sec:theory}
\subsection{Irwin-Hall error analysis}\label{subsec:IrwinHallError}
Define the random variable 
\begin{equation}
    Z_n=\sqrt{12/n}(X_n-n/2).
    \label{eq:Z_n}
\end{equation}
Let $f_{Z_n}$ be the pdf of $Z_n$ and $f_\mathcal{N}$ be the pdf of $\mathcal{N}(0,1)$.
\begin{theorem}\label{thm: IrwinHallError}
$\|f_{Z_n}-f_{\mathcal{N}}\|_{\infty}\leq O\left(\frac{1}{n}\right)$.
\end{theorem}
The proof is given in Appendix.

\begin{figure}[h]
    \centering
    {\bf a)} \includegraphics[width=3.2in]{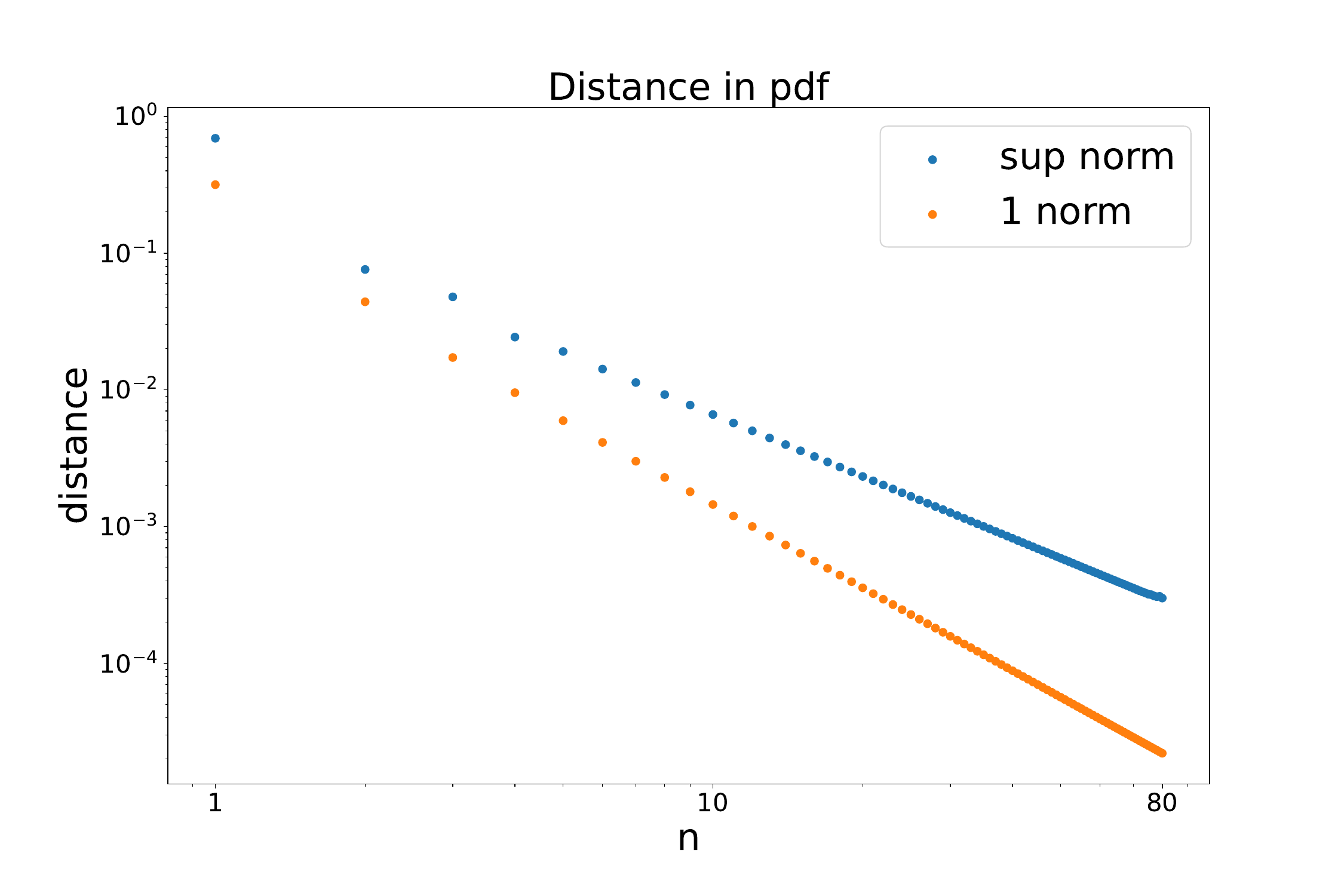}
    {\bf b)} \includegraphics[width=3.2in]{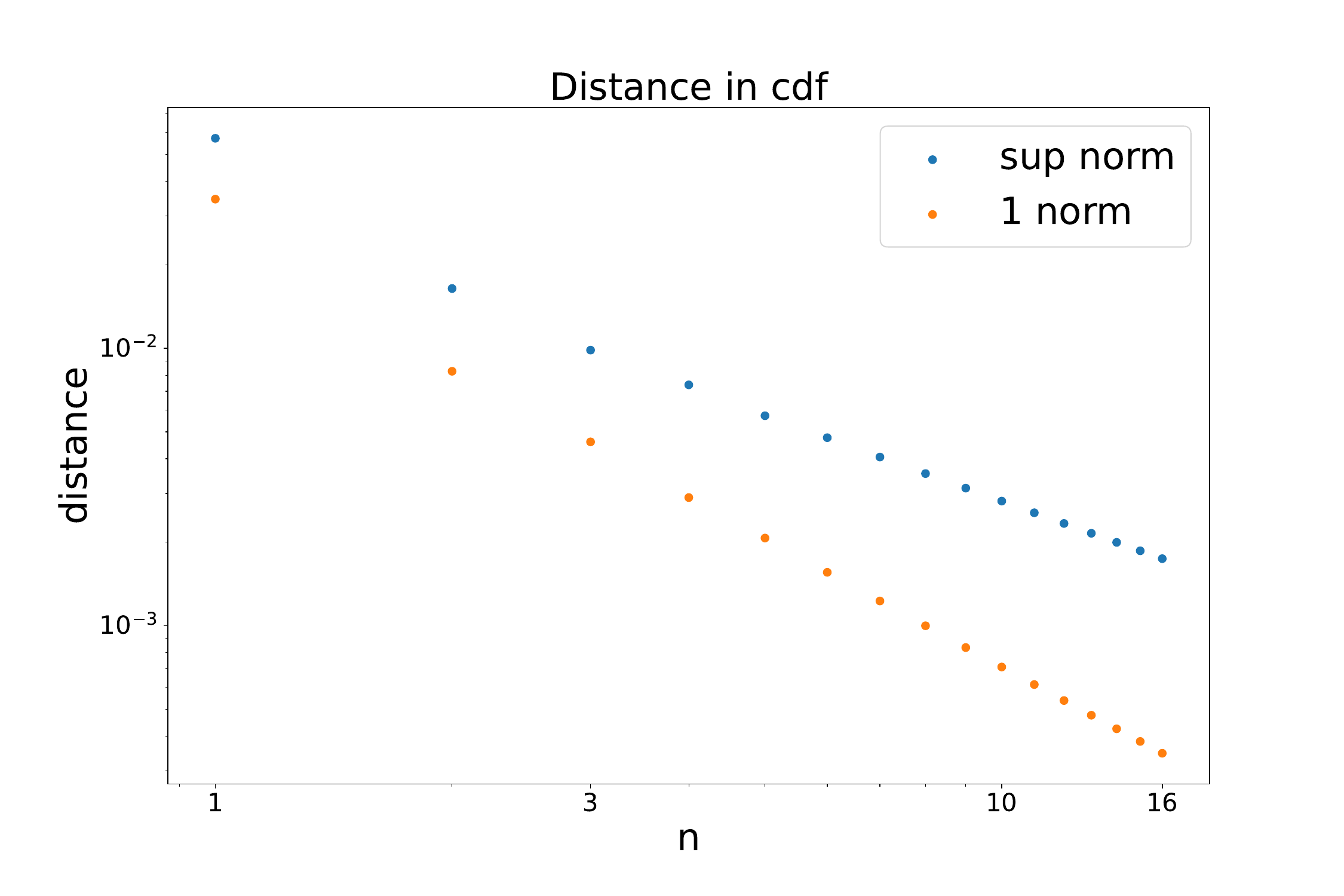}
    \caption{{\bf Distance between Irwin-Hall distributions and normal distributions.} {\bf (a)} Distance in pdf between Irwin-Hall distribution and normal distribution as the Irwin-Hall number $n$ grows. {\bf (b)} Distance in cdf between Irwin-Hall distribution and normal distribution as the Irwin-Hall number $n$ grows.}
    \label{fig:cdf_irwinhall}

\end{figure}

The error between pdfs of Irwin-Hall distributions and the corresponding normal distribution is also analyzed numerically. Since $f_{Z_n}$ has bounded support on $[-\sqrt{3n},\sqrt{3n}]$, and $f_\mathcal{N}(x)\to 0$ monotonically as $|x|\to \infty$, it is sufficient to consider $\|f_{Z_n}-f_{\mathcal{N}}\|_{\infty}$ on $[-\sqrt{3n},\sqrt{3n}]$. We take points with spacing $\frac{\sqrt{3}}{50\sqrt{n}}$ in this domain and numerically evaluate $f_{Z_n}$ and $f_\mathcal{N}$ on these points. We numerically estimate the gradient of the max error to be -1.57 and that of the average error to be -2.06. Therefore, based on numerical calculation shown in Fig.~\ref{fig:cdf_irwinhall} a),
\begin{align}
    \|f_{Z_n}-f_{\mathcal{N}}\|_{\infty}\approx\Theta\left(\frac{1}{n^{1.57}}\right),\\
    \|f_{Z_n}-f_{\mathcal{N}}\|_1\approx\Theta\left(\frac{1}{n^{2.06}}\right).\label{eq:pdf1norm}
\end{align}

We also list similar results for the cdf.
\begin{lemma}\cite{Sherman1971}
$\|F_{Z_n}-F_{\mathcal{N}}\|_\infty\leq O\left(\frac{1}{n}\right)$.
\end{lemma}
The error between cdf of Irwin-Hall distributions and the corresponding normal distribution is analyzed numerically as well. Based on numerical calculation show in Fig.~\ref{fig:cdf_irwinhall} b),
\begin{align}
    \|F_{Z_n}-F_{\mathcal{N}}\|_{\infty}\approx\Theta\left(\frac{1}{n^{1.18}}\right),\\
    \|F_{Z_n}-F_{\mathcal{N}}\|_1\approx\Theta\left(\frac{1}{n^{1.61}}\right).
\end{align}
We also compare the theoretical error scaling of our method against other methods. In Ref. \citeonline{Rattew2021efficient}, it was numerically studied that the KL divergence between the distribution produced by the algorithm and the corresponding exact analytical normal distribution scales as $\Theta(1/t^2)$, where $t$ counts the iterations and can be thought of as proportional to the circuit depth. Since $\log(p/q)\approx (p-q)/q$ for $p\approx q$, the 1-norm between the distribution produced by Ref. \citeonline{Rattew2021efficient} and the corresponding exact analytical normal distribution also scales as $\Theta(1/t^2)$. However, this algorithm uses the Mid-Circuit Measurement and Reuse scheme, and is thus not suitable for implementation with quantum Monte Carlo algorithms \cite{Montanaro2015}.

\subsection{Discretization error}\label{subsec:discretizationerror}
\begin{figure}[h]
    \centering
    \includegraphics[width=4in]{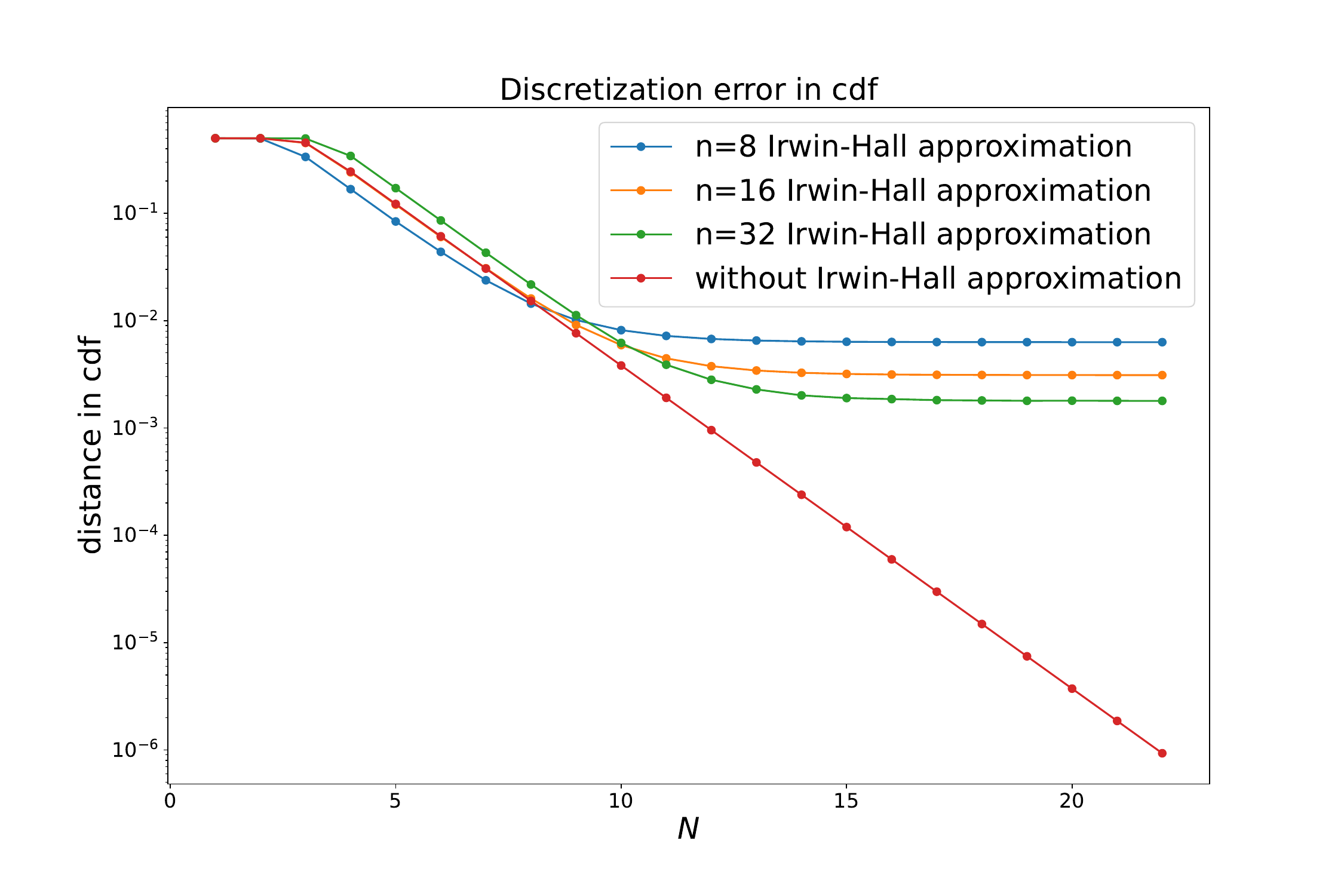}
    \caption{{\bf Error (sup norm) in cdf when comparing discretized Irwin-Hall and discretized normal distribution pdf to the undiscretized normal distribution.}}
    \label{fig:discretizationerror}
\end{figure}

We also discuss briefly the discretization error. The normal distribution is a continuous probability distribution. However, in modern computers, decimal numbers are typically represented as as 32/64/128 bit floats and there is some inherent discretization. When the distribution is loaded onto a quantum state, there is further discretization due to the finite number of qubits. As an example, let us assume the Irwin-Hall order of 16. We discretize $\mathcal{N}(0, 1)$ and $Z_n$ (Irwin-Hall of order $n$) for $n=16$ over the domain of $[-4\sqrt{3},4\sqrt{3}]$ over qubits up to 23, and compared against the corresponding $\mathcal{N}(0, 1)$ over $(-\infty,\infty)$ in Fig. \ref{fig:discretizationerror}.

As we can see, the error from the discretized Irwin-Hall distribution and from the discretized normal coincides for less than 9 qubits. But for the discretized Irwin-Hall distribution, the error plateaus after 12-13 qubits, indicating that the error now mostly comes from the Irwin-Hall approximation to the normal distribution and not the discretization.

\subsection{MPS Circuit Approximation Error Analysis}\label{sec:mpserror}
\begin{figure}[h]
\centering
{\bf a)}
\includegraphics[width=3.2in]{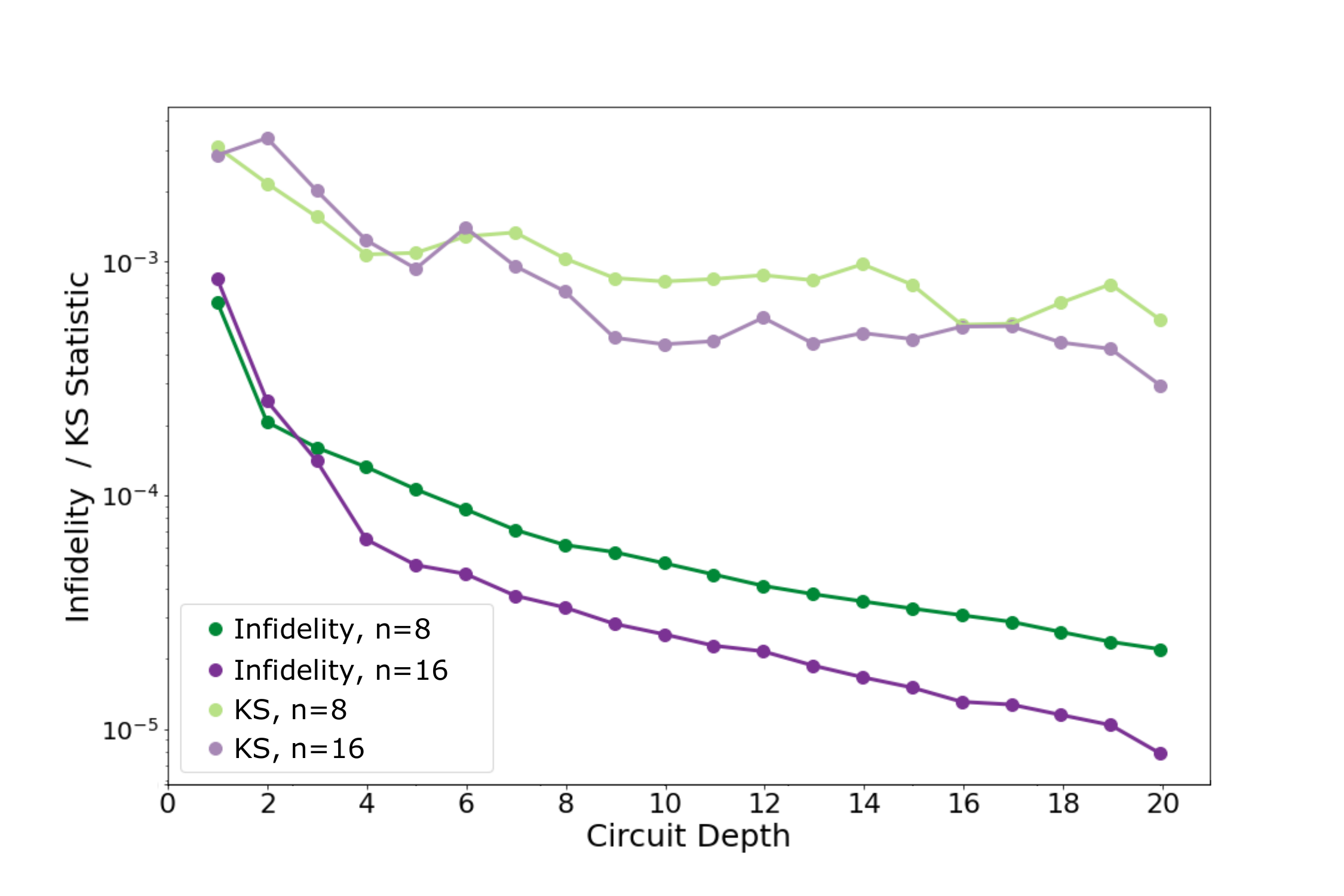}
{\bf b)}
\includegraphics[width=3.2in]{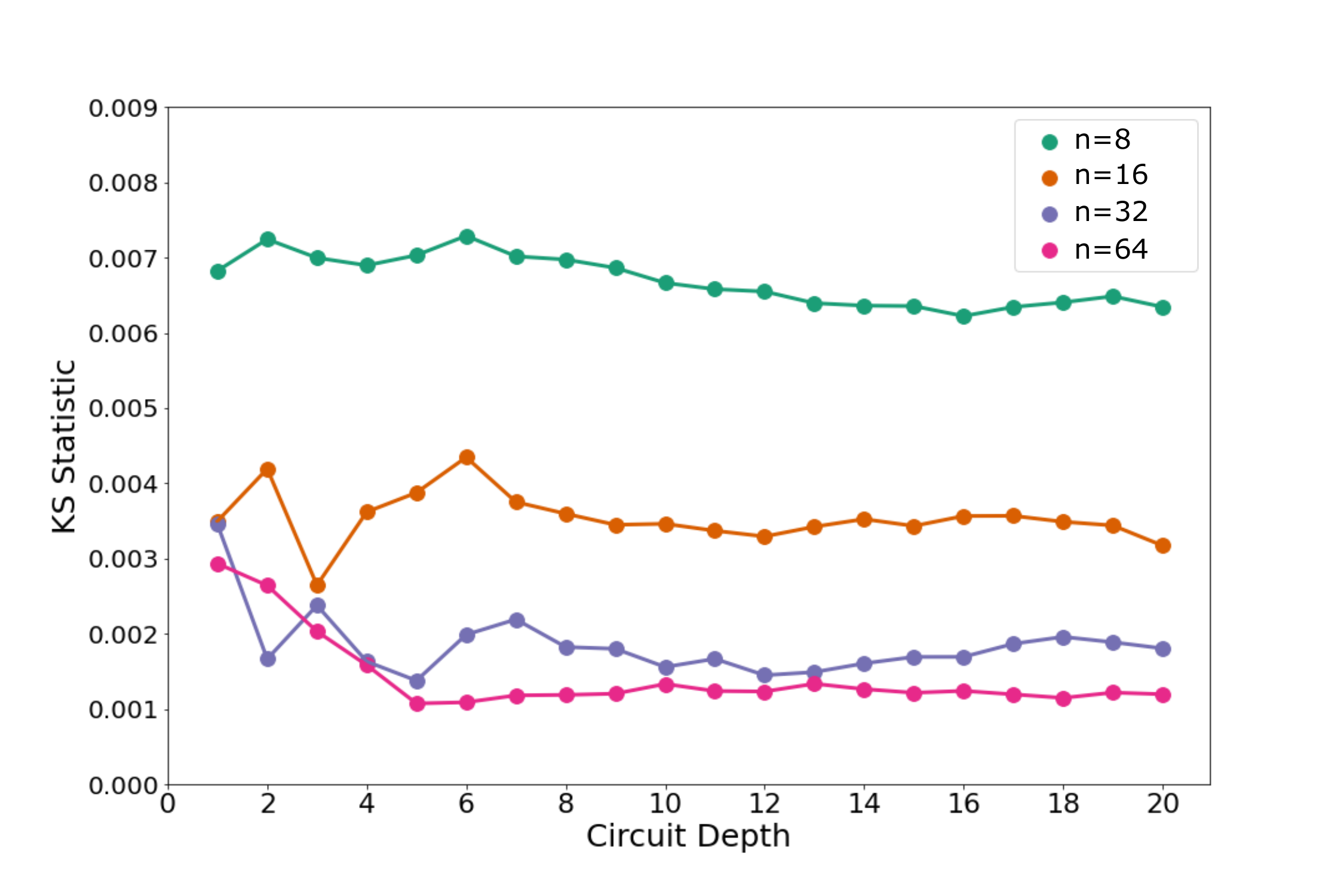}
\caption{{\bf The error introduced by the matrix product state circuit when approximating the target ideal normal distribution.} {\bf a)} The error between the distribution of the prepared wave function and target Irwin-Hall
    distribution as the number of MPS layers increases. We show both the
    wave function infidelity and the KS statistic between the output probability
    distributions. {\bf b)} The KS statistic between the prepared distributions
    and the ideal normal distributions for different Irwin-Hall orders as a
    function of MPS circuit depth. These plots are based on simulations of the quantum circuits with $N=14$ qubits.}
\label{fig:ks_mps_D}
\end{figure}

We also look at the convergence of our iterative MPS loading scheme to the ideal normal distribution as circuit depth is increased. First, we make a distinction between the target Irwin-Hall distribution, which is the state we load into the untruncated MPS wave function, and the ideal normal distribution which is the ultimate target distribution. As the number of layers in the
iterative MPS circuit construction is increased, the approximation error between the output of the quantum circuit, $\ket{\psi}$, and target Irwin-Hall distribution $\ket{\phi}$, strictly decreases. This behavior can be seen in Fig.~\ref{fig:ks_mps_D} a), where we prepare Irwin-Hall distributions with $n = 8$ and $16$. We find that the infidelity, $I=1-|\langle\psi|\phi\rangle|$, empirically decays as $I\sim\left(\frac{1}{D}\right)^\alpha$ for large D with $\alpha = 1.08$ for $n=8$ and $\alpha \approx 1.22$ for $n=16$. We also fit the average decay rate for all $n$ in Appendix.
We note that in both cases, the infidelity appears to decay more rapidly for the first few layers before slowing down as more layers are added.  Also, it has recently been proposed that a gate-by-gate optimization method discussed in Ref. \citeonline{zapata_mps}
may lead to a more rapid decay of the infidelity, especially as the number of layers in the MPS circuits is increased. 

We also look at the KS statistic between output and target distributions, which
measures the maximum difference between the cumulative density functions of the
two distributions. This metric has additional experimental relevance as it can be directly
estimated from a finite number of measurements taken in the computational basis
on the prepared quantum state.  In Fig.~\ref{fig:ks_mps_D} a), we show the
behavior of the KS statistic between $\ket{\psi}$ and the target Irwin-Hall
distribution as the number of MPS layers increases.   Again, we see that the
general trend also decreases as a power law with increasing layer depth. However, in this case
the behavior of KS statistic is not a smooth function and is not strictly decreasing.
This simply demonstrates that state fidelity is not always in one-to-one
correspondence with all measures of distance between the prepared and target
output distributions.

This fact is further emphasized when we compare distributions prepared by our
quantum circuit simulation and the ideal normal distribution. We plot this comparison in
Fig.~\ref{fig:ks_mps_D} b), for $n=8,16,32$ and $64$ and $N=14$. At high circuit depth, where we apply many circuit layers of the iterative MPS state preparation procedure, we very closely reproduce the exact Irwin-Hall distribution in the amplitudes of the quantum state. 
We find that in all
cases, a single layer of the MPS circuit  already achieves a fairly low value
for the KS-statistic, comparable with the error originating solely from the
Irwin-Hall approximation as shown in Fig.~\ref{fig:discretizationerror}. We also find that differences
between the target Irwin-Hall distribution and the target ideal distribution in
many cases offset the differences between the prepared distribution and the target Irwin-Hall distribution. This results in relatively strong fluctuations in the KS
statistic value with increasing circuit depth. In all cases, the Irwin-Hall
error dominates the MPS approximation error beyond depth $D=5$. In fact, it is
only for $n=32$ and $n=64$, that there appears to be any significant
benefit in going beyond $D=1$.  

In summary, in this section we studied the error introduced to our state
preparation routine by the Irwin-Hall approximation, qubit discretization and the MPS circuit
preparation routine. We find that for the Irwin Hall orders which we consider, a circuit with gates linear in the number of qubits $N$ and with only a small number of layers $D$, is
sufficient to prepare a probability distribution with error comparable to that
introduced by the Irwin Hall approximation. We believe these values of $n$, $N$ and $D$ will be sufficient for most practical applications for the near future. Of course, a more accurate
approximation of the normal distribution will require a higher circuit depth
to prepare.  In the next subsection we provide an analysis of how these quantities must scale in this higher precision limit. However, any easy-to-prepare probability distribution which
approximates the exact normal distribution will likely be well approximated by
a finite depth MPS circuit.

\subsection{Resource Estimations}\label{sec:resource}
\begin{figure}[h]
    \centering
    \includegraphics[width=4in]{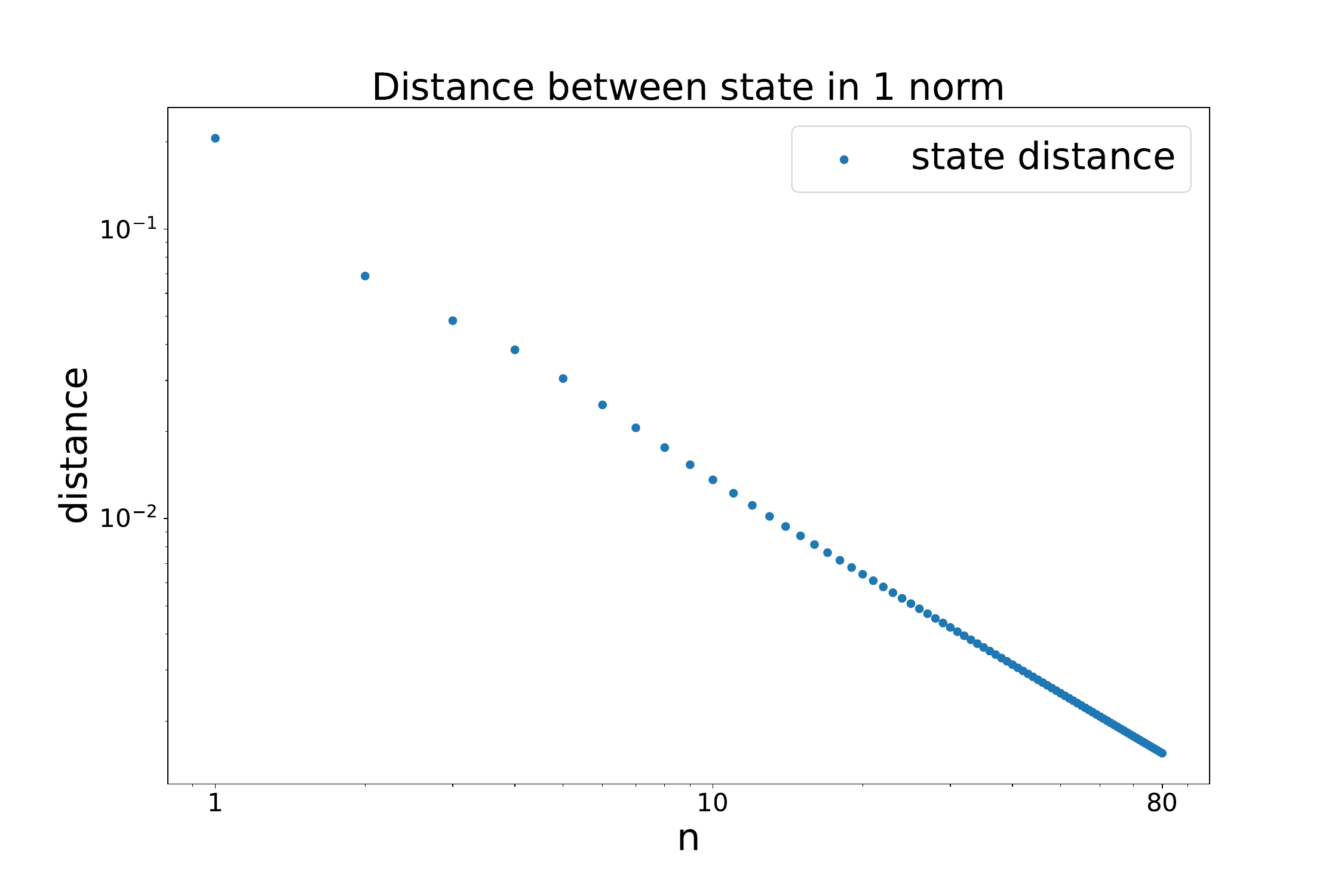}
    \caption{\bf Distance between the ideal state and Irwin-Hall approximated state in L2 norm ($\epsilon$ as defined in Eqn. \ref{eq:epsilon})}
    \label{fig:state_distance}
\end{figure}

Here we list the computational resources of our algorithm, both from classical and quantum point of view in table \ref{tab:resource}, with respect to $\epsilon$, as defined in Eq.~\ref{eq:epsilon} as $||\ket{\psi} - \ket{\tilde{\psi}}|| \le \epsilon$, where $\ket{\psi}$ is the wave functions with amplitudes proportional to the square root of ideal normal distribution and $\ket{\tilde{\psi}}$ our MPS state. Note that we assume both the Irwin-Hall approximation and the MPS approximation introduces error of $O(\epsilon)$.
\begin{table}[h]
\begin{center}
    \begin{tabular}{|c|c|c|}
         \hline
         Irwin-Hall Computation & MPS Approximation & Quantum Circuit \\
         \hline
         $\Tilde{O}(\epsilon^{-1.87})$& $\Tilde{O}(\epsilon^{-3.74})$& $\Tilde{O}(\epsilon^{-1.74} )$\\
         \hline
    \end{tabular}\caption{Computational complexities of different components of the algorithm with respect to $\epsilon$}\label{tab:resource}
\end{center}
\end{table}
First, we numerically study the relation between $\epsilon$ and the Irwin-Hall order $n$ as shown in Fig. \ref{fig:state_distance}.
We derive $\epsilon=O(n^{-1.07})$, and therefore $n=O(\epsilon^{-1/1.07})$. In Appendix, we showed that classically computing all the coefficients of an Irwin-Hall distribution of order $n$ has complexity $O(n^2)=O(\epsilon^{-1.87})$.

For an Irwin-Hall distribution of order $n$, our MPS is a sum of $n$ piece-wise MPS components each with bond dimension $n+1$. Each piece-wise MPS component involves the calculation of $Nn^2$ coefficients so that the total classical cost of this calculation is $\Tilde{O}(Nn^3)$. The sum of these component is then a MPS with  $\chi=n(n+1)$, and the cost of generating this MPS state is $N \chi^2 \sim N n^4$, which is the dominant contribution in this portion of the calculation. This scaling can be reduced by truncating the MPS bond dimension $\chi$ after each addition of the piece-wise MPS components which each has $\chi_p=(n+1)$, instead of only truncating after summing all components. In this case, there are $n$ computations each with complexity $\chi_p^2=n^2$, leading to an overall complexity of $Nn^3$.  The classical processing cost associated with truncating the MPS to bond dimension $\chi'<\chi$ is $N \chi'^3$ using the variational compression scheme and $N\chi^2\chi'$ using the SVD compression method \cite{Schollwock_2011}. When generating the quantum circuit, we must repeatedly disentangle the MPS state $D$ times, then truncate to bond-dimension $\chi'=2$. This has complexity $\Tilde{O}(N \chi'^2 D)$. Therefore overall, the cost associated with generating the quantum circuit for preparing an Irwin-Hall distribution of order $n$ is $\Tilde{O}(N n^4 + N + ND)$, where we ignore the constant cost associated with $\chi'$. We saw that $n=O(\epsilon^{-1/1.07})$. We also saw that the discretization error decays exponentially with the number of qubits $N$, so that $\epsilon \sim e^{-\gamma N}$, for some constant $\gamma$, so that $N \sim \log(1/\epsilon)$. In Appendix,
we show that $\epsilon^2 \sim 1/D^b$ for some power $b$, which we numerically estimate to be $b\ge 1.15$ for the Irwin-Hall functions we studied in this work. Therefore we can write $D \sim \epsilon^{-1.74}$, so that the classical processing of the MPS manipulations scales as $\Tilde{O}( \log(1/\epsilon) [\epsilon^{-4/1.07} + \epsilon^{-1.74}] )=\Tilde{O}(\epsilon^{-3.74})$ after dropping polylog terms and subleading terms.

The quantum circuit corresponding to $D$ rounds of MPS approximations has circuit depth $O(DN)$. In Appendix, we show that for a given $\epsilon$, the error associated with the MPS circuit approximation to the exact Irwin-Hall wave function is essentially independent of both the number of qubits $N$ and the Irwin-Hall order $n$. Furthermore, we show that $\epsilon$ decays like a power of the circuit depth $D$, so that $\epsilon^2 \sim 1/D^b$ for some constant $b$. We numerically show that $b\ge 1.15$. Therefore the quantum circuit complexity is given by $\Tilde{O}(ND) = \Tilde{O}(\log(1/\epsilon) \epsilon^{-1.74})=\Tilde{O}(\epsilon^{-1.74})$.

We expect it is possible to improve the complexity scaling of both the classical MPS approximation and the MPS circuit depth $D$. For example, if we truncate the MPS bond dimension $\chi$ after each addition of the piece-wise MPS components, we will get the MPS approximation to scale as $\Tilde{O}(\epsilon^{-2.80})$ instead of $\Tilde{O}(\epsilon^{-3.74})$. We believe it is possible to improve the MPS circuit depth $D$ through better approximation methods for low depth circuit construction, for example using Ref. \citeonline{zapata_mps}. In this sense, the exponent $b$ in the scaling $\Tilde{O}(ND) = \Tilde{O}(\epsilon^{-2/b})$ should be thought of as an upper bound for this analysis. Furthermore, we see in Appendix, that if we were to use the exact circuit construction for an MPS of bond dimension $\chi'$, instead of the low-depth circuit approximation, the error appears to decrease exponentially as $\epsilon \sim e^{-b \chi'}$ for some constant $b$. The depth of an exact MPS circuit construction scales like $D \sim \Tilde{O}(\chi'^2)$, which would imply that the circuit complexity would scale like $\Tilde{O}(\log^2(1/\epsilon))$.  While the constant overhead of the exact MPS circuit implementation implies it is less practical at moderate values of $\epsilon$ than the low-depth approximations, it may provide a large advantage when trying to prepare states with very small approximation error $\epsilon$.

Note that normal distributions of different parameters can all be obtained from the standard one by rescaling and shifting. So parameters of the normal distribution do not contribute to the circuit complexity here.

\subsection{Applicability to Monte Carlo Integration}\label{subsec:MC}
The resource estimation of our algorithm allows an end-to-end error analysis of applications where our algorithm can be a subroutine. As an example we discuss in detail how our algorithm would affect the results of a Monte Carlo integration.

As a reminder we assume $||\ket{\psi} - \ket{\tilde{\psi}}|| \le \epsilon$, where $\ket{\psi}$ is the wave functions with amplitudes proportional to the square root of ideal normal distribution and $\ket{\tilde{\psi}}$ our MPS state. We also implicitly assume both the Irwin-Hall approximation and the MPS approximation introduces error of $O(\epsilon)$. 

In classical Monte Carlo integration, we estimate $\mathbb{E} [g]$ by $\frac{1}{s}\sum_{i=0}^{s}g(x_i)$, with $x_i$ sampled from the desired distribution. The central limit theorem states that
\begin{equation}
    \frac{1}{s}\sum_{i=0}^{s}g(x_i)\to \mathcal{N}(\mathbb{E}[g],\sigma/\sqrt{s}),
\end{equation}
where $s$ is the number of times that we sample the data.
So the error of Monte Carlo integration is $O(\sigma/\sqrt{s})$, where $\sigma^2=\text{Var}[g]$.

This assumes $x_i$ is sampled from the desired distribution. In our case, we are using the Irwin-Hall distribution together with MPS to approximate $\mathcal{N}$, so it will introduce additional error as $g(x_i)$'s are biased estimators now. The additional error is $O(\epsilon)$, as we show in Appendix.

We may also express the complexities of classical and quantum Monte Carlo integration in terms of $\delta$, which is the accuracy that we want a Monte Carlo integration to achieve. Classically, assuming sampling from a normal distribution takes $O(1)$, the complexity of a Monte Carlo simulation is proportional to the number of samples $s$, which is $O(\delta^{-2})$. We note here that our assumption on the sampling complexity of normal distribution is fairly strong and unlikely to be true. More analysis on the error dependence of classical sampling algorithms for normal distributions, such as the Box-Muller transform\cite{Box1958} or Ziggurat algorithm\cite{Marsaglia1961}, is needed.

In quantum Monte Carlo integration, we have $t=\Tilde{O}(\delta^{-1})$. Assuming the additive error introduced by the state preparation algorithm is comparable to $\delta$, we have $O(\epsilon)=\delta$. Thus we can express the computational resources of our algorithm in terms of $\delta$ as
\begin{table}[h]
\begin{center}
    \begin{tabular}{|c|c|c|}
         \hline
         Irwin-Hall Computation & MPS Approximation & Quantum Circuit \\
         \hline
         $\Tilde{O}(\delta^{-1.87})$& $\Tilde{O}(\delta^{-3.74})$& $\Tilde{O}(\delta^{-1.74} )$\\
         \hline
    \end{tabular}\caption{Computational complexities of different components of the algorithm with respect to $\delta$}\label{tab:montecarlo}
\end{center}
\end{table}

The end-to-end complexity of a quantum Monte Carlo algorithm leveraging our state preparation subroutine will be $\Tilde{O}(\delta^{-2.74})$, as the state preparation subroutine is used $t$ times. The one-time classical pre-processing of our state preparation subroutine is dominated by the MPS approximation, which scales as $\Tilde{O}(\delta^{-3.74})$. In reality, it's possible to complete the classical pre-processing before-hand, and have the quantum subroutine as part of a standard state preparation toolkit.

As seen from this analysis (table \ref{tab:montecarlo}), quantum Monte Carlo integration using our current implementation of state preparation does not provide an asymptotic advantage over classical Monte Carlo integration. However, as pointed in Section \ref{sec:resource}, exact MPS constructions can be used in the low $\delta$ limit, which should improve the complexity.

\subsection{Experimental Demonstration}\label{subsec:exp}
\begin{figure}[h]
\centering
{\bf a)}
\includegraphics[height=1.51in]{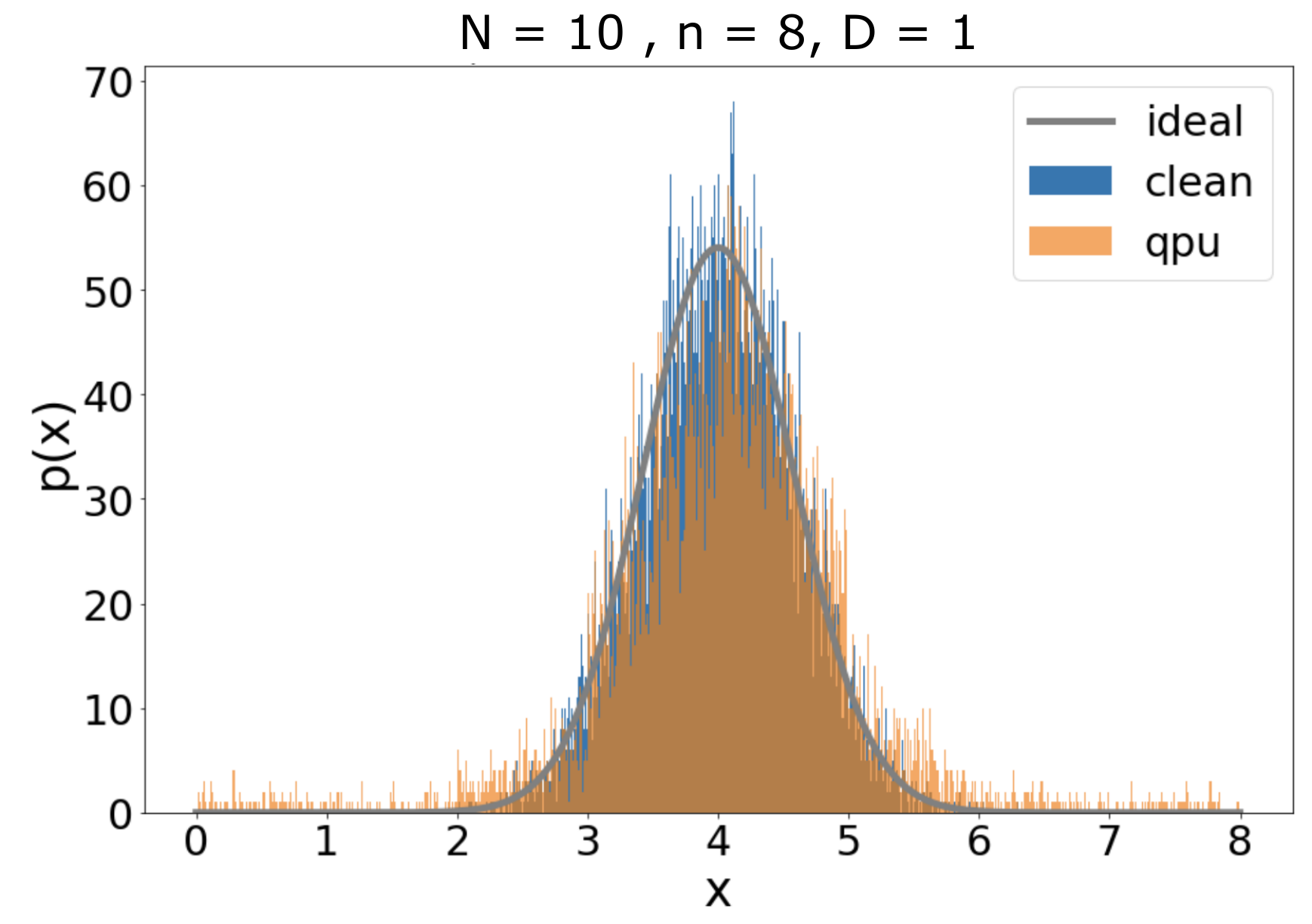}
{\bf b)}\includegraphics[height=1.51in]{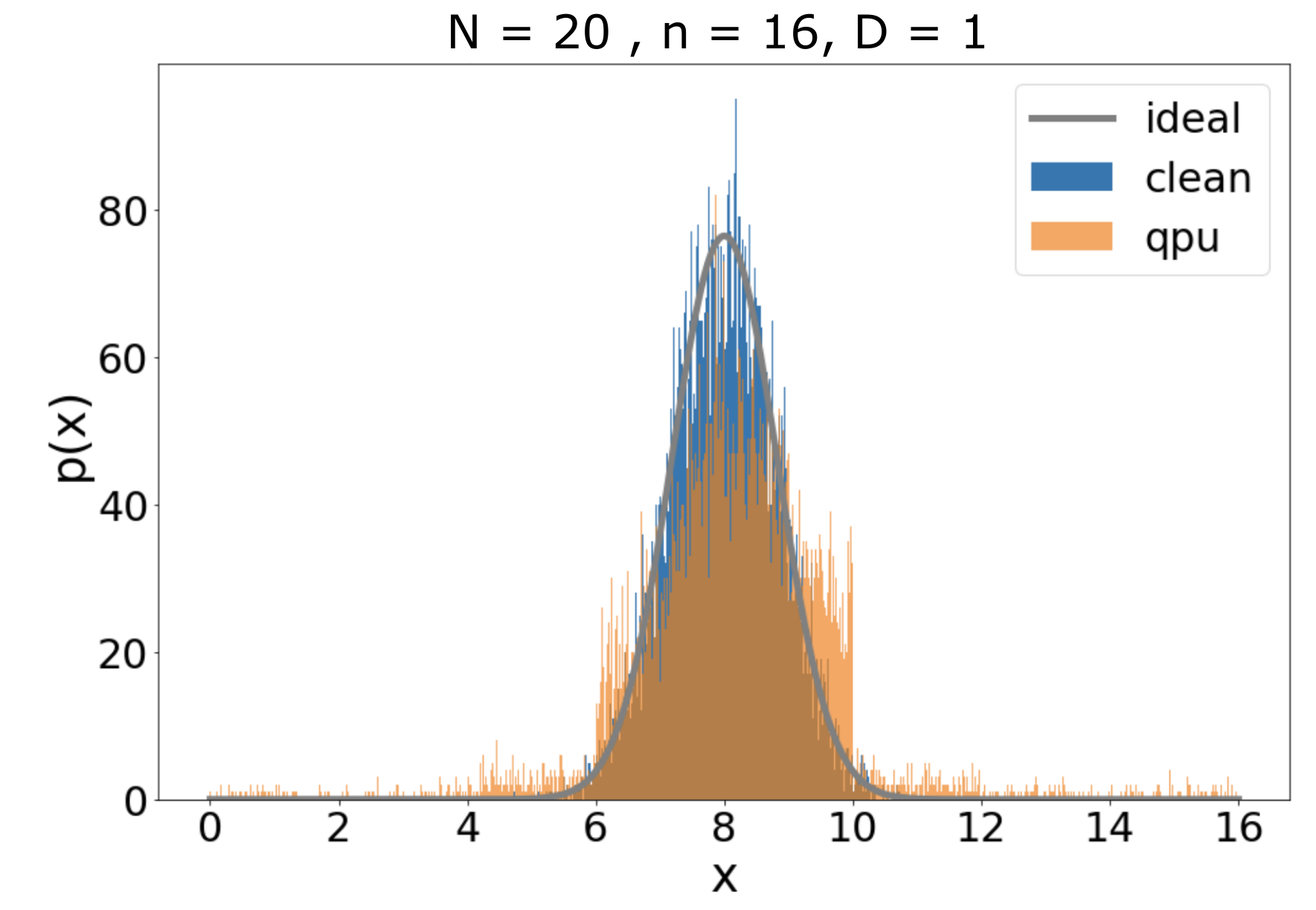}
{\bf c)}\includegraphics[height=1.51in]{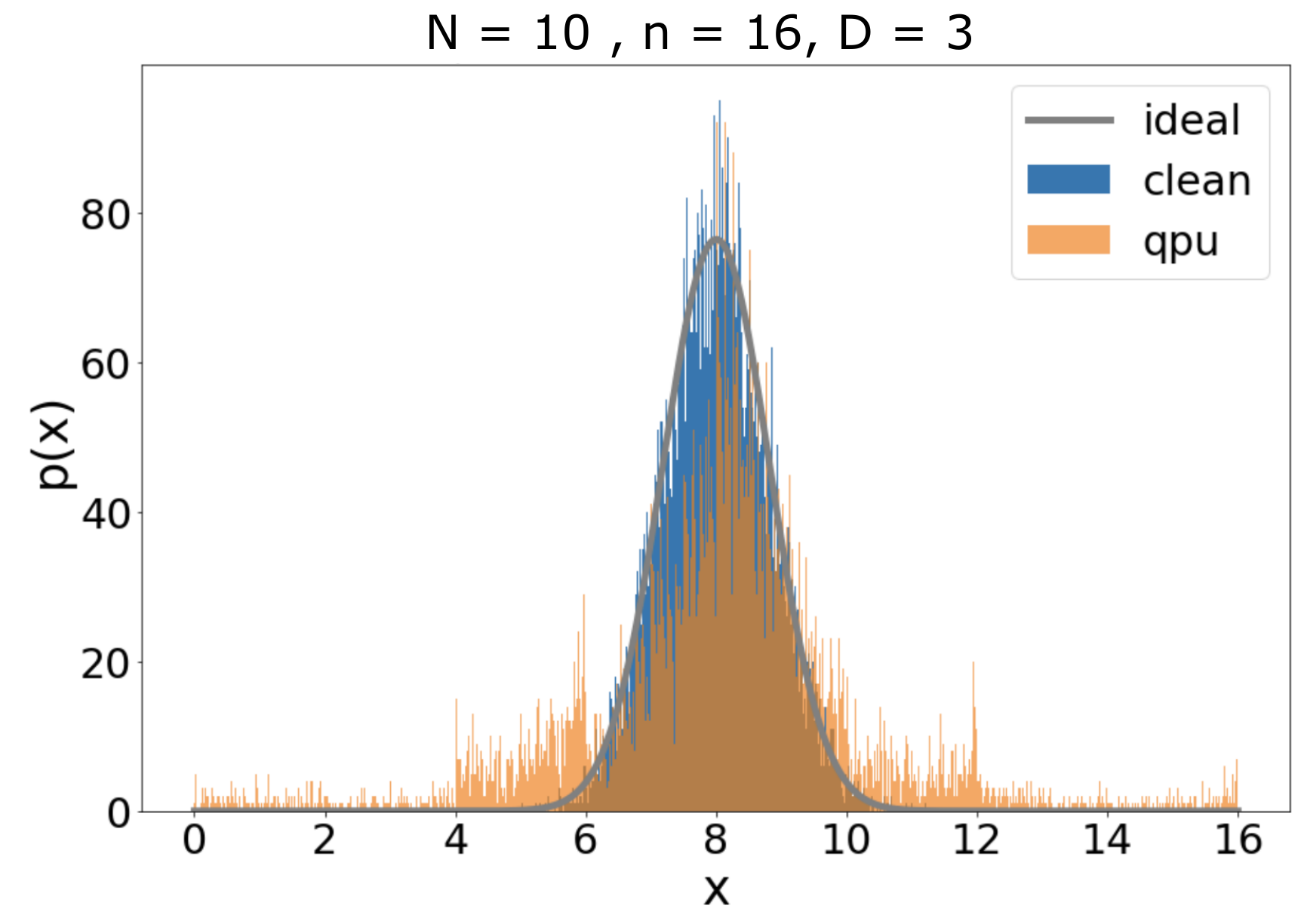}
\caption{{\bf Histogram of measurements for normal distributions which were prepared on the quantum computer.} Experiments were run on IonQ Aria on circuits with between 10 to 20 qubits, with MPS depth D between 1 to 3. Each circuit uses $2\times (N-1)\times D$ CX gates and 10000 shots were taken on each circuit. Here we show three illustrative implementations, with (a) $N=8, n=8, D=1$, (b) $N=20, n=16, D=1$ and (c) $N=10,n=16,D=3$. In all figures, samples taken from the QPU are shown in orange, and can be compared to an equal number of samples drawn from the noiseless simulator in blue, and to the ideal pdf shown in gray. }
\label{fig:exp_hist}
\end{figure}

In this section, we show the experimental demonstration of our state preparation procedure on the IonQ Aria generation of trapped ion quantum computer. 
First, in Fig.~\ref{fig:exp_hist}, we demonstrate our ability to prepare normal probability distributions with different variances on quantum states with 10 to 20 qubits. This involves preparing all $2^{10}$ to $2^{20}$ amplitudes of the quantum wave function. By using the MPS state preparation method, all amplitudes can be set by applying only between 20-120 CNOT gates. We also highlight that to the best of our knowledge, this is the largest demonstration of state preparation for applications in quantum Monte Carlo to date. 

Demonstration of the full algorithm on the quantum hardware requires compiling the generic 4x4 unitary gates generated by the iterative MPS encoding procedure into a set of fundamental gates, which we choose to be arbitrary single qubit rotations plus the 2-qubit CNOT gate.  Note that when running on the IonQ hardware, each CNOT gate is automatically transpiled to a single native 2-qubit 
Molmer-Sorensen gate, so that the total number of basic 2-qubit gates remains the same. Also note that the output of the MPS encoding are real 4x4 unitary matrices, implying that each gate is an $O(4)$ transformation. We therefore use the result of Ref.~\citeonline{vatan_williams}, in order to compile each unitary to 2 CNOT gates plus a potential SWAP gate, as show in Fig.~\ref{fig:mps_circuit}. On the IonQ architecture, the all-to-all connectivity of the lattice implies that SWAP gates can be removed at the cost of a reordering of the qubits in software. Therefore, each $O(4)$ transformation is implemented using only 2 CNOT gates. 

We show three illustrative examples of the output distributions of the prepared wave functions. For each set of hyper-parameters, we make 10000 measurements and plot the histogram of the measured bit strings, converted to the scale of the original Irwin-Hall distribution $X_n$ (So that $\mu=n/2)$. The standard deviation of the distributions is $\sigma = \sqrt{n/24}$. (Note, while $X_n$ is an approximation to $\mathcal{N}(n/12,\sqrt{n/12})$, we are comparing to $\mathcal{N}(n/12, \sqrt{n/24})$. This is because we are preparing quantum state with probabilities proportional to the pdf of $X_n$, and this corresponds to a quantum state with probability amplitudes the square root of the pdf of $X_n$. Because the square root of the pdf of a normal distribution $\mathcal{N}(\mu,\sigma)$ is the pdf of $\mathcal{N}(\mu,\sigma/\sqrt{2})$, and the fact that $X_n\to \mathcal{N}(n/12,\sqrt{n/12})$, the standard deviation of the distribution is $\sqrt{n/24}$.)  We see that the experimental measurements give good visual agreement with the clean simulation and the ideal pdf in all cases. As the quantum gate noise is the largest source of error in the algorithm, we find that the best results are always obtained for a single layer of the MPS iterative preparation procedure. In general, the fewer gates applied the higher the fidelity of the output distribution. We can visualize this behavior in Fig.~\ref{fig:exp_hist}. The shortest depth circuits we apply prepare 10-qubit systems with MPS depth 1. We show the case when $n=8$, and see very strong agreement with the expected histogram on 10000 samples using the noiseless simulator. The effect of the noise is to slightly broaden the normal peak and to create additional samples at the tails of the distribution.  The same general behavior is seen when preparing a 20 qubit system with MPS depth 1. We show the case when $n=16$. In this case, the number of basic gates is roughly doubled, resulting in a higher rate of noise and a broader measured output distribution.  However, we still see a strong visual overlap with the simulated noiseless distribution and the ideal pdf function.  Finally, in order to better visualize the effects of noise, we also show the output for preparing a 10 qubit system at depth 3. We see that in this case, the noise is significantly higher.

\begin{figure}[h]
\centering
{\bf a)}
\includegraphics[width=3.2in]{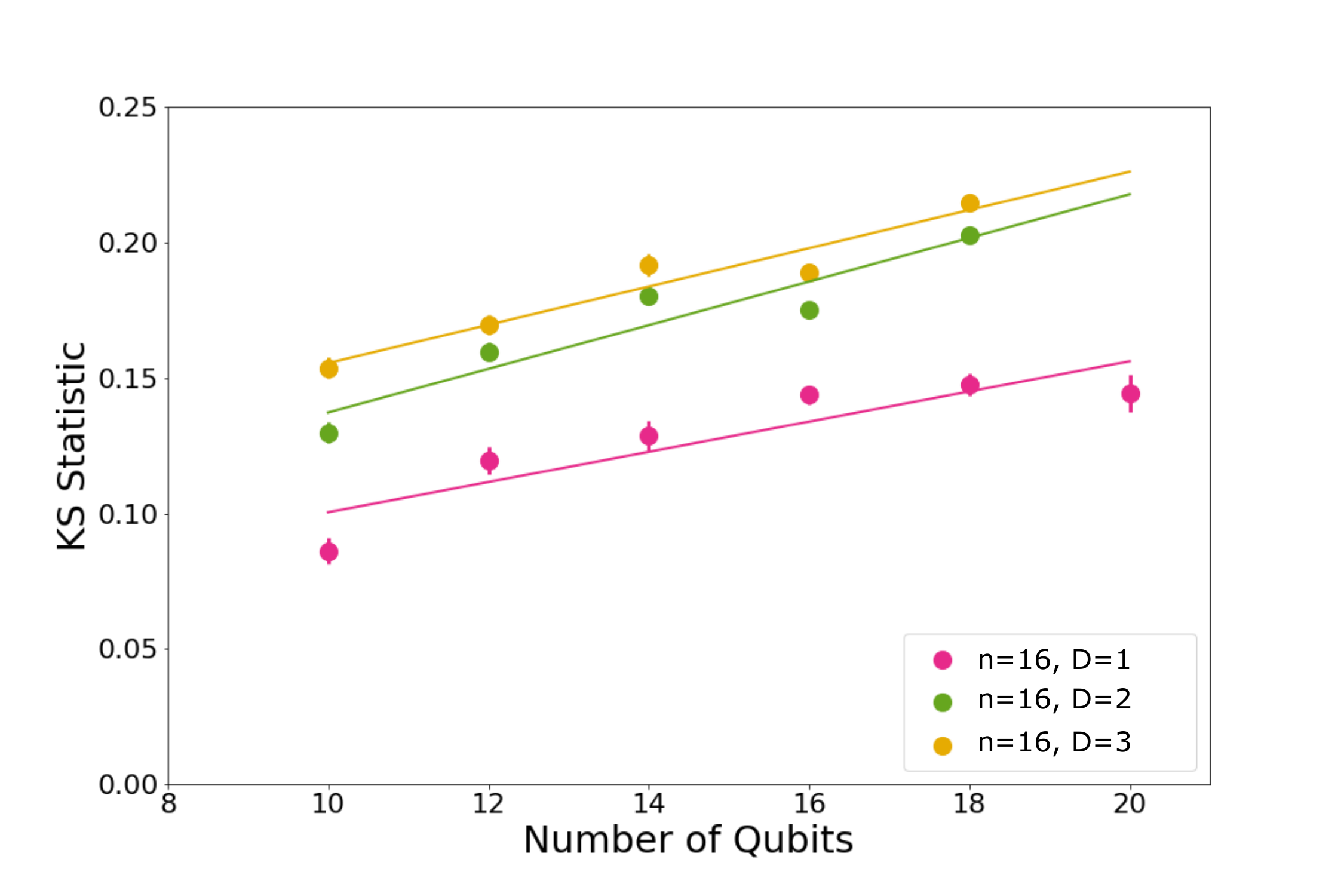}
{\bf b)}
\includegraphics[width=3.2in]{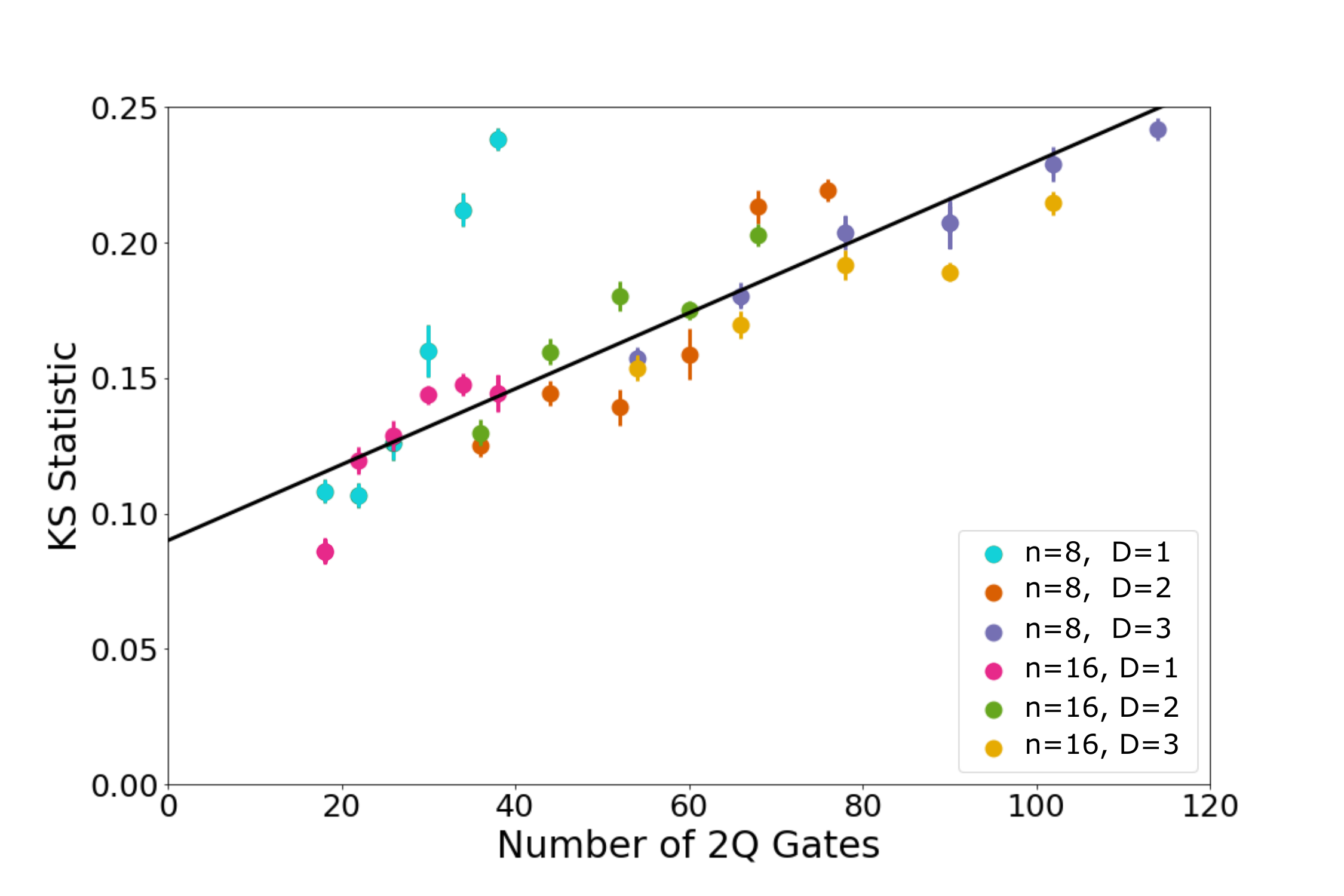}
\caption{{\bf The KS Statistic between the ideal distribution and the distribution sampled from the experimental implementation.} {\bf a) } Scaling of the KS statistic with circuit width (Number of Qubits) for MPS depths $D=1,2,3$ and $n=16$. As explained in the text, with current device noise levels, optimal state preparation occurs with $D=1$. {\bf b)} Scaling of the KS statistic with number of two-qubit gates demonstrates that these circuit operations are the dominant source of noise in the state preparation procedure.}
\label{fig:exp_ks}
\end{figure}

We now quantify the effect of noise over the full set of prepared normal distributions. We show the results in Fig.~\ref{fig:exp_ks}. We evaluate the quality of our experimental state preparation by comparing the finite number of measurements from the quantum circuit to measurements sampled from an ideal normal distribution using the Kolmogorov–Smirnov statistic.  In  Fig.~\ref{fig:exp_ks}a) we show the KS statistic for the $n=16$ normal distributions for 10-20 qubits and 1-3 MPS layers.   As mentioned, the best results are found for a single MPS layer. The best KS statistic value we find is $D_n \approx 0.09$, and increases to around $D_n\approx 0.15$ for the largest system sizes. We can compare these values to the value required to pass the KS test of the equality of the measured and target distributions given by Eq.~\ref{eq:KS_test}. Using this equation, we can expect to pass this test with statistical significance $\alpha=0.05$, when the number of samples, $s$, taken from both distributions is $s\leq 450$ for $D_n=0.09$, and $s\leq 150$ for $D_n=0.15$.  In contrast to the noiseless case, on the noisy quantum computer, the value of the KS statistic increases with number of layers and number of qubits. That is, the noise generated from the additional gates overcomes the higher fidelity of the underlying normal approximation when preparing states using higher number of MPS layers.  There is a clear increase in the KS statistic both as the number of qubits and the number of MPS layers increases. In general, the largest source of noise on near term quantum computers comes from the execution of two-qubit gates. The overall effect of this noise can be clearly seen in Fig.~\ref{fig:exp_ks} b), where we plot KS statistic vs the total number of two-qubit gates in the associated quantum circuit. While there can be significant variance in the noise from circuit to circuit, likely originating from small fluctuations in system performance with time, the overall trend is consistent. The KS statistic ranges from 0.09 to 0.25 in the worst case, and appears to increase linearly with the total number of two-qubit gates in the circuit. 

Note that while we did not extend this pattern down to zero gates, the y-intercept is expected to be nonzero due to the presence of state-preparation and measurement (SPAM) errors. While we expect SPAM errors to be relatively low on ion trap quantum computers, they still likely make a nontrivial contribution to the overall error. Also, we cannot reliable apply our state preparation procedure in the zero gate limit and so the approximation error may be a large contributor to the overall error in this limit. Therefore the zero-gate extrapolation error is likely a combination of both SPAM errors and the state preparation approximation error.

\section{Discussion} \label{sec:discussion}
Our paper gives a procedure that produces $U$ such that $U\ket{0^{N}}$ has probability amplitudes proportional to a normal distribution. Our circuit for $U$ does not involve any ancilla qubit and succeeds with certainty. This has a number of advantages over other methods in applications where $U$ needs to be applied iteratively \cite{Montanaro2015}. If an algorithm only applies $U$ with probability $1-\delta$ \cite{Rattew2021efficient,Rattew2022}, then $m$ sequential applications of $U$ would succeed with probability $(1-\delta)^{m}$, which may be not sufficient for a practical application. If an algorithm uses $k$ ancillary qubits \cite{Rattew2021efficient,Zhang2022,Rattew2022} such that the ancillary qubits cannot be discarded or reused between successive applications of $U$, then $m$ sequential applications of $U$ would need $km$ ancillary qubits, which is undesirable since quantum computers will be resource-constrained in the foreseeable future.


Our Matrix Product State based procedure, on the other hand, uses only $N$ qubits to prepare a discretized probability distribution on $2^{N}$ points, and succeeds with probability 1 in the limit of noiseless quantum gates. The trade-off for these advantages is that we can only approximately prepare the distribution to fixed accuracy for a quantum circuit with gates linear in the number of qubits. The main sources of this error originate from approximating our target function with a piece-wise polynomial function and from approximately preparing the low bond dimension MPS wave function with a short depth quantum circuit. 

Throughout this work, we carefully analyzed the error originating from both these sources.  We chose to approximate the normal distribution using the order $n$ Irwin-Hall distribution, which gives a fast, deterministic method for generating the appropriate unitary circuit $U$. We found that the error from this approximation decreases polynomially with $n$.  In order to approximately encode this Irwin-Hall distribution into the wave function amplitudes using a short depth circuit, we use the iterative circuit construction method of Ref.~\citeonline{ran2020encoding}. While our method does not give the best theoretical complexity in terms of $\epsilon$ (as defined in Section \ref{subsec:statepreparation}), the short circuit depth makes it far more appealing than other methods with better asymptotic complexity. Also, intrinsic restrictions on hardware accuracy and discretization errors arising from floating-point arithmetic mean that it may not be possible to go down to infinitesimal $\epsilon$ in a practical application.

We also note that the circuit structure of our algorithm, as depicted in Fig. \ref{fig:mps_circuit}, only assumes nearest-neighbour interaction of the underlying device. This implies that our algorithm can be implemented on a wide range of hardware architecture, such as ion trap devices, superconducting devices and so on, without any additional swap gate.

We experimentally validated this loading technique on a trapped ion quantum computer for a range of circuits with varying width and depth.  We were able to successfully generate the target wave functions on circuits with up to 20 qubits. Furthermore, the measured KS statistics were comparable to previous state preparation experiments \cite{Zoufal2019}, while acting on circuits with many more qubits. We also studied at the scaling of the KS statistic with circuit depth, and found that, as expected, with the hardware noise is currently the largest source of error in this state preparation procedure, limiting the effectiveness of applying higher depth loading circuits.  With our current experiments, we are approaching the capacity of generating single precision random quantum variables, although we expect a large improvement in gate-error rates is still required to take full advantage of such high precision numbers. 

A number of possible directions stemming from this work immediately present themselves for future research.  One possibility is to study improvements to the MPS approximation method for low depth circuit construction. Another is to extend this work to study more general probability distributions. This involves searching for simple piece-wise polynomial representations of these distributions with controllable approximation errors. Finally, we leave to future research a full end-to-end implementation of a quantum amplitude estimation algorithms which makes use of this state preparation subroutine, and an associated analysis of the error rates required to achieve quantum advantage using these approaches. 

\section{DATA AVAILABILITY}
The data supporting the study’s findings are available from the corresponding author upon reasonable request.
\section{CODE AVAILABILITY}
The code supporting the study’s findings are available from the corresponding author upon reasonable request.
\section{ACKNOWLEDGEMENT}
This work is a collaboration between Fidelity Center for Applied Technology, Fidelity Labs, LLC., and IonQ Inc. The Fidelity publishing approval number for this paper is 1075809.2.0.
\section{AUTHOR CONTRIBUTIONS}
J.I. proposed using matrix product states for state preparation, designed the quantum circuits, performed experiments, and carried out the error analysis. S.J. contributed to the design of the algorithm and experimental error analysis. E.Z. initiated the idea of using Irwin-Hall polynomials, and contributed to theoretical error analysis.
\section{COMPETING INTERESTS}
The authors declare no competing interests.

\clearpage
\newpage
\appendix
\renewcommand{\figurename}{Supplementary Figure}
\setcounter{equation}{0}
\setcounter{figure}{0}
\section*{Supplementary Note}
\subsection*{Convergence of the Irwin-Hall distribution}\label{appdix:IrwinHall}
Here we present the proof of Theorem \ref{thm: IrwinHallError} in the main text.
\begin{proof}
Let $\{U_1,\cdots,U_n\}$ be $n$ i.i.d. uniform random variable $U(-1/2,1/2)$. Then $f_{U_i}(x)=1$ if $|x|\leq 1/2$ and 0 otherwise. The characteristic function of $U_i$ is
\begin{equation}
    \varphi_{U_i}(t)=\frac{\sin t}{t}.
\end{equation}

Define the random variable
\begin{equation}
    Z_n=\frac{\sqrt{3}\sum_{i=1}^n U_i}{\sqrt{n}}.
\end{equation}
Then the characteristic function of $Z_n$ is given by
\begin{equation}
    \varphi_{Z_n}(t)=\Pi_{i=1}^n \varphi_{U_i}(\sqrt{3}t/\sqrt{n}).
\end{equation}

By properties of the characteristic function, the probability density function of $Z_N$ is given by
\begin{align}
    f_{Z_n}(x)&=\frac{1}{2\pi}\int_{-\infty}^\infty e^{-itx}\varphi_{Z_n}(t) dt \\
    &=\frac{1}{2\pi}\int_{-\infty}^\infty\cos (tx)\left(\frac{\sin(\sqrt{3}t/\sqrt{n})}{\sqrt{3}t/\sqrt{n}}\right)^n dt
\end{align}

Similarly, the probability density function of the standard normal distribution $N(0,1)$ can be written as
\begin{equation}
    f_N(x)=\frac{1}{2\pi}\int_{-\infty}^\infty\cos (tx)e^{-t^2/2} dt
\end{equation}

Therefore
\begin{align}
    \sup_x|f_{Z_n}(x)-f_N(x)|&=\frac{1}{2\pi}\sup_x\left|\int_{-\infty}^\infty\cos (tx)\left(\left(\frac{\sin(\sqrt{3}t/\sqrt{n})}{\sqrt{3}t/\sqrt{n}}\right)^n-e^{-t^2/2}\right)dt\right|\\
    &\leq \frac{1}{\pi}\left|\int_{0}^\infty\left(\left(\frac{\sin(\sqrt{3}t/\sqrt{n})}{\sqrt{3}t/\sqrt{n}}\right)^n-e^{-t^2/2}\right)dt\right|\\
    &\leq \frac{1}{\pi}\left|\int_{0}^{\frac{\pi}{2}\sqrt{\frac{n}{3}}}\left(\left(\frac{\sin(\sqrt{3}t/\sqrt{n})}{\sqrt{3}t/\sqrt{n}}\right)^n-e^{-t^2/2}\right)dt\right|\\
    &+\frac{1}{\pi}\left|\int_{\frac{\pi}{2}\sqrt{\frac{n}{3}}}^\infty\left(\frac{\sin(\sqrt{3}t/\sqrt{n})}{\sqrt{3}t/\sqrt{n}}\right)^n dt\right|+\frac{1}{\pi}\int_{\frac{\pi}{2}\sqrt{\frac{n}{3}}}^\infty e^{-t^2/2}dt.
\end{align}
Here in the second line we used the fact that the integral is symmetric in $t$ and $|\cos(tx)|\leq 1$ and in the third and fourth line we used the triangle inequality.

Examining the upper bound by pieces, we see that
\begin{align}
    &\frac{1}{\pi}\left|\int_{0}^{\frac{\pi}{2}\sqrt{\frac{n}{3}}}\left(\left(\frac{\sin(\sqrt{3}t/\sqrt{n})}{\sqrt{3}t/\sqrt{n}}\right)^n-e^{-t^2/2}\right)dt\right|\\
    =&\frac{1}{\pi}\sqrt{\frac{n}{3}}\left|\int_{0}^\frac{\pi}{2}\left(\left(\frac{\sin(t)}{t}\right)^n-e^{-nt^2/6}\right)dt\right|
\end{align}

Using the inequality
\begin{equation}
    e^{-t^2/6-t^4/138}\leq \sin(t)/t\leq e^{-t^2/6}~~~~~~~~~~~~\text{for}~~0 \leq t\leq \frac{\pi}{2},
\end{equation}
we get an upper bound
\begin{align}
    &\frac{1}{\pi}\sqrt{\frac{n}{3}}\int_0^{\frac{\pi}{2}}\left(e^{-nt^2/6}\left(1-e^{-nt^4/138}\right)\right) dt\\
    \leq &\frac{1}{\pi}\sqrt{\frac{n}{3}}\int_0^{\frac{\pi}{2}}\left(\frac{nt^4}{138}e^{-nt^2/6}\right) dt\\
    =& \frac{1}{138\sqrt{3}\pi n}\int_0^{\frac{\pi}{2}\sqrt{n}}t^4e^{-t^2/6}dt=O\left(\frac{1}{n}\right),
\end{align}
where in the second line we used the inequality $1-e^{-x}\leq x$ and in the third line we used integration by substitution.

The second piece in the upper bound can be upper bounded as
\begin{align}
    &\frac{1}{\pi}\left|\int_{\frac{\pi}{2}\sqrt{\frac{n}{3}}}^\infty\left(\frac{\sin(\sqrt{3}t/\sqrt{n})}{\sqrt{3}t/\sqrt{n}}\right)^n dt\right|\\
    \leq &\frac{1}{\pi}\int_{\frac{\pi}{2}\sqrt{\frac{n}{3}}}^\infty\left(\frac{1}{\sqrt{3}t/\sqrt{n}}\right)^n dt=\exp\left(-O(n)\right).
\end{align}

The third piece in the upper bound can be upper bounded by $\exp(-O(n))$, as
\begin{equation}
    \text{erfc}(x)\leq \exp(-x^2).
\end{equation}
Therefore
\begin{equation}
    \sup_x|f_{Z_n}(x)-f_N(x)|\leq O\left(\frac{1}{n}\right)
\end{equation}
\end{proof}

\subsection*{Time Complexity Analysis of Irwin-Hall Coefficients Computation}\label{appdix:irwinhallcomplexity}

In this appendix, we present detailed analysis for the computational complexity of Irwin-Hall coefficients.

Restricting to $x\in [k,k+1]$, Eqn. \ref{eq:irwinhallpdf} from the main text can be rewritten as
\begin{equation}
    f_{X_n}(x)=\frac{1}{(n-1)!}\sum_{j=0}^{n-1}a_j(k,n)x^j,
\end{equation}
where the coefficients $a_j(k,n)$ can be computed from a recurrence relation over $k$
\begin{equation}
    a_j(k,n)=
    \begin{cases}
    1\hspace{2em} &k=0,j=n-1\\
    0\hspace{2em} &k=0,j<n-1\\
    a_j(k-1,n)+(-1)^{n+k-j-1}\binom{n}{k}\binom{n-1}{j}k^{n-j-1}\hspace{2em} &k>0.
    \end{cases}    
\end{equation}
To obtain all the coefficients, we can first compute $(n-1)!$, $\binom{n}{k}$, $\binom{n-1}{j}$ for all $k,j$. A naive algorithm will compute $(n-1)!$ with complexity $O(n)$. Using the dynamic programming algorithm to calculate all $\binom{n}{k}$, $\binom{n-1}{j}$ requires time complexity of $O(n^2)$ and space complexity of $O(n^2)$. Next we can compute $k^{n-j-1}$ for all $k,j$ with time complexity $O(n^2)$ and space complexity $O(n^2)$. With all them on hand, we can compute $a_j(k,n)$ for all $k,j$ using the above recurrence relation with dynamic programming, with an additional time complexity of $O(n^2)$ and space complexity of $O(n^2)$. Adding everything up gives a time complexity of $O(n^2)$ for computing all the Irwin-Hall coefficients of order $n$.

\subsection*{Error Analysis of MPS Approximation}\label{appdix:error}

In this appendix, we present some more detailed analysis of the error associated with approximating the exact MPS representation of the Irwin Hall distribution with a truncated MPS approximation, and study the error associated with approximating this MPS representation with a state prepared using the iterative quantum circuit procedure described in the main text.  In both cases, we compare the error in term of the infidelity $ I=1-|\langle \tilde{\psi}|\psi\rangle|\leq\frac{1}{2}\epsilon^2$, between the approximate MPS state and the exact Irwin Hall MPS state, so that we ignore any contributions to the error which arise from the Irwin Hall approximation.

First we show that our state preparation procedure at fixed depth results in a similar infidelity, nearly independent of the number of qubits  in the quantum state. This favorable scaling with $N$ one of the major advantages of this method, which is theoretically justified by the entanglement arguments of Section \ref{sec:MPS} in the main text, and which implies that we can prepare normal distribution to high precision with low depth circuits In Supplementary Fig.~\ref{fig:Nscaling}, we plot the infidelity between the target quantum state (which encodes the target Irwin Hall distribution at order n) and the approximate MPS state at depth $D$ is approximately constant with the number of qubits between 8 and 20 at depths $D=1,2$ and $3$, with only minor fluctuations.
\begin{figure}[H]
\centering
\includegraphics[width=2.5in]{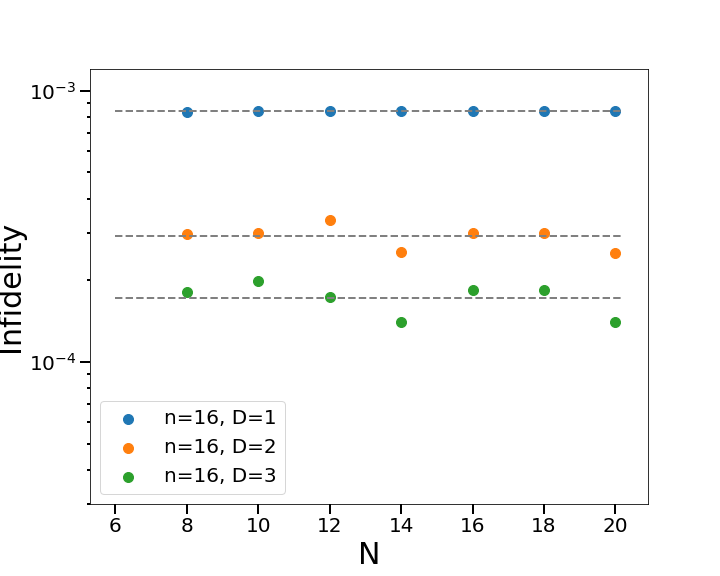}   
\hspace{8mm} \includegraphics[width=2.5in]{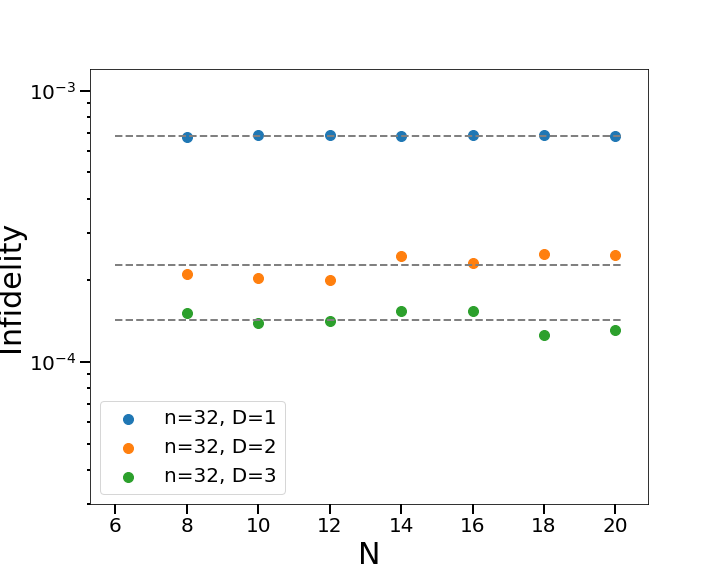}
\caption{The error between the state generated by the iterative quantum circuit with $D$ layers and the exact Irwin Hall matrix product state with $n=16$ {\it(left)} and $n=32${\it(right)}, as a function of the number of qubits. We see that the infidelity is largely independent of $N$}
\label{fig:Nscaling}
\end{figure}

We also show that this infidelity is largely independent of the Irwin Hall order $n$.
In Supplementary Fig.~\ref{fig:nih_scaling} see that at fixed circuit depth $D$, the infidelity is approximately independent of the Irwin Hall order n. 
\begin{figure}[H]
\centering
\includegraphics[width=2.5in]{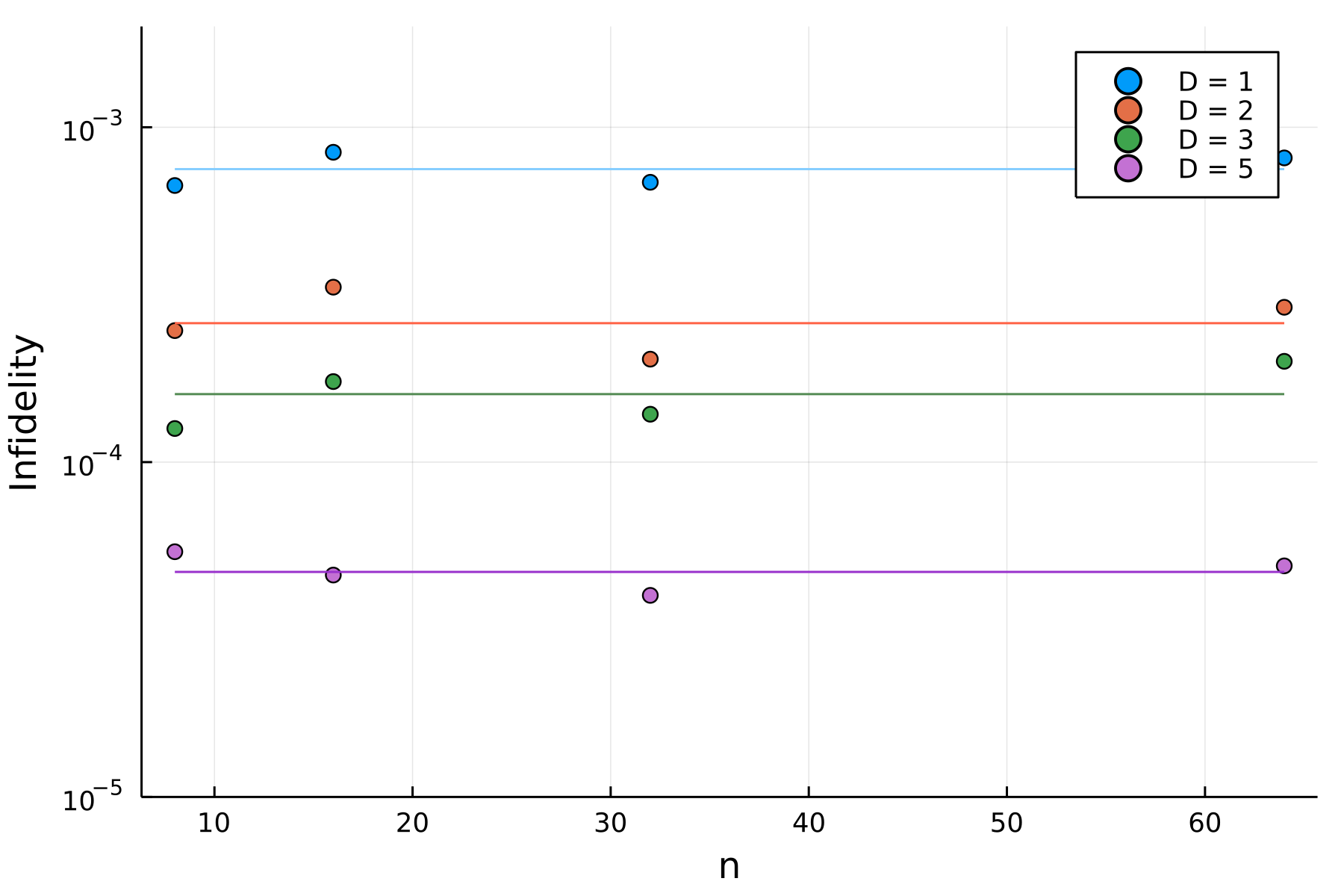}
\caption{Infidelity between the state generated by the iterative quantum circuit with $D$ layers and the exact Irwin Hall matrix product state with $N=12$ qubits as a function of the Irwin Hall order $n$.}
\label{fig:nih_scaling}
\end{figure}

Therefore, although the discretization error associated to the number of qubits $N$ and the error associated with applying the Irwin Hall approximation both contribute to the overall error, these factors largely do not affect the quality of the MPS approximation error.

There also exist two sources of error which are introduced by the MPS approximation to the exact wave function, and the quantum circuit approximation to the MPS state. The MPS approximation error depends only on the truncated bond dimension $\chi'$, and appears to decreases exponentially with the bond dimension $\chi'$.

\begin{figure}[H]
\centering
\includegraphics[width=2.8in]{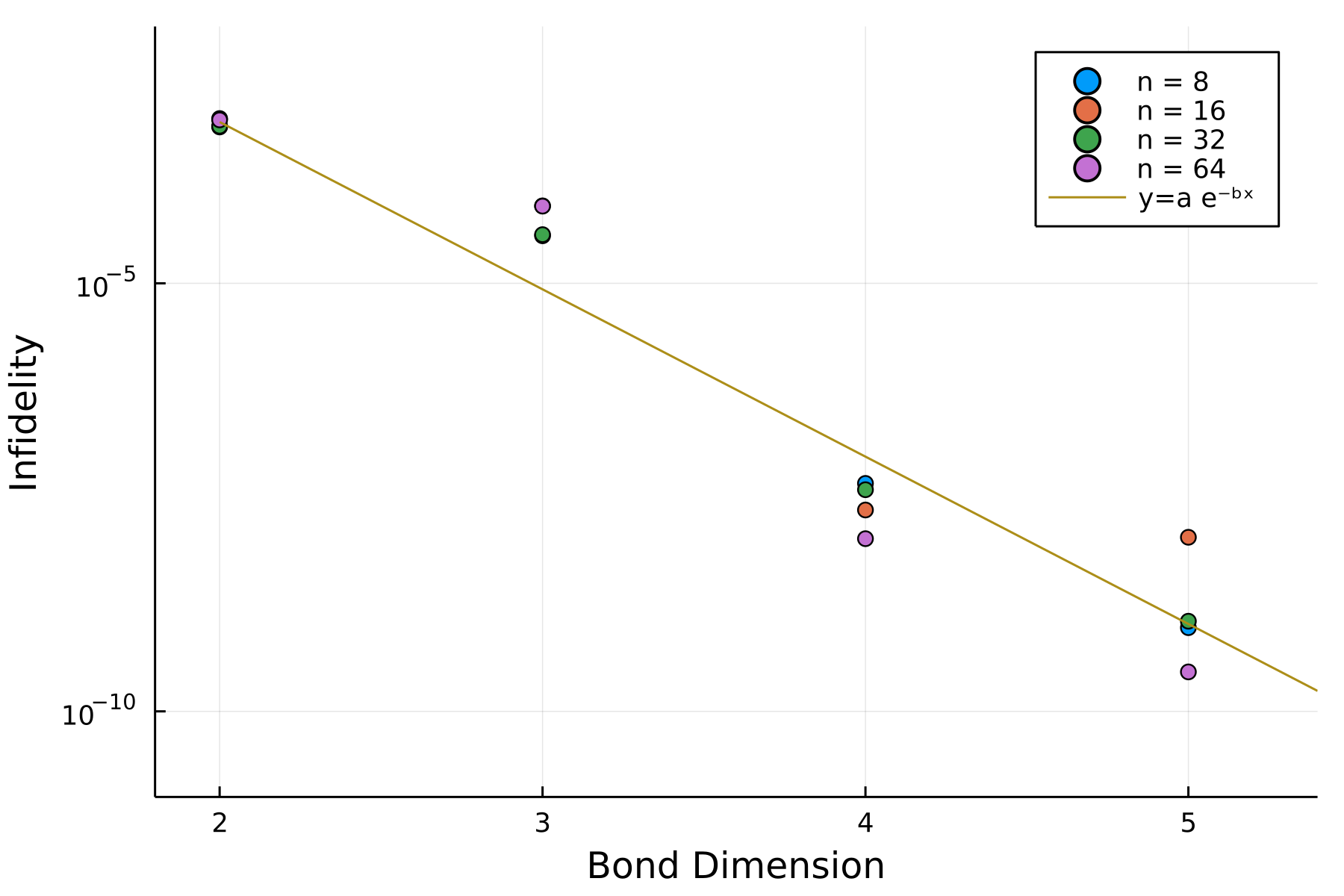}   
\hspace{8mm} \includegraphics[width=2.8in]{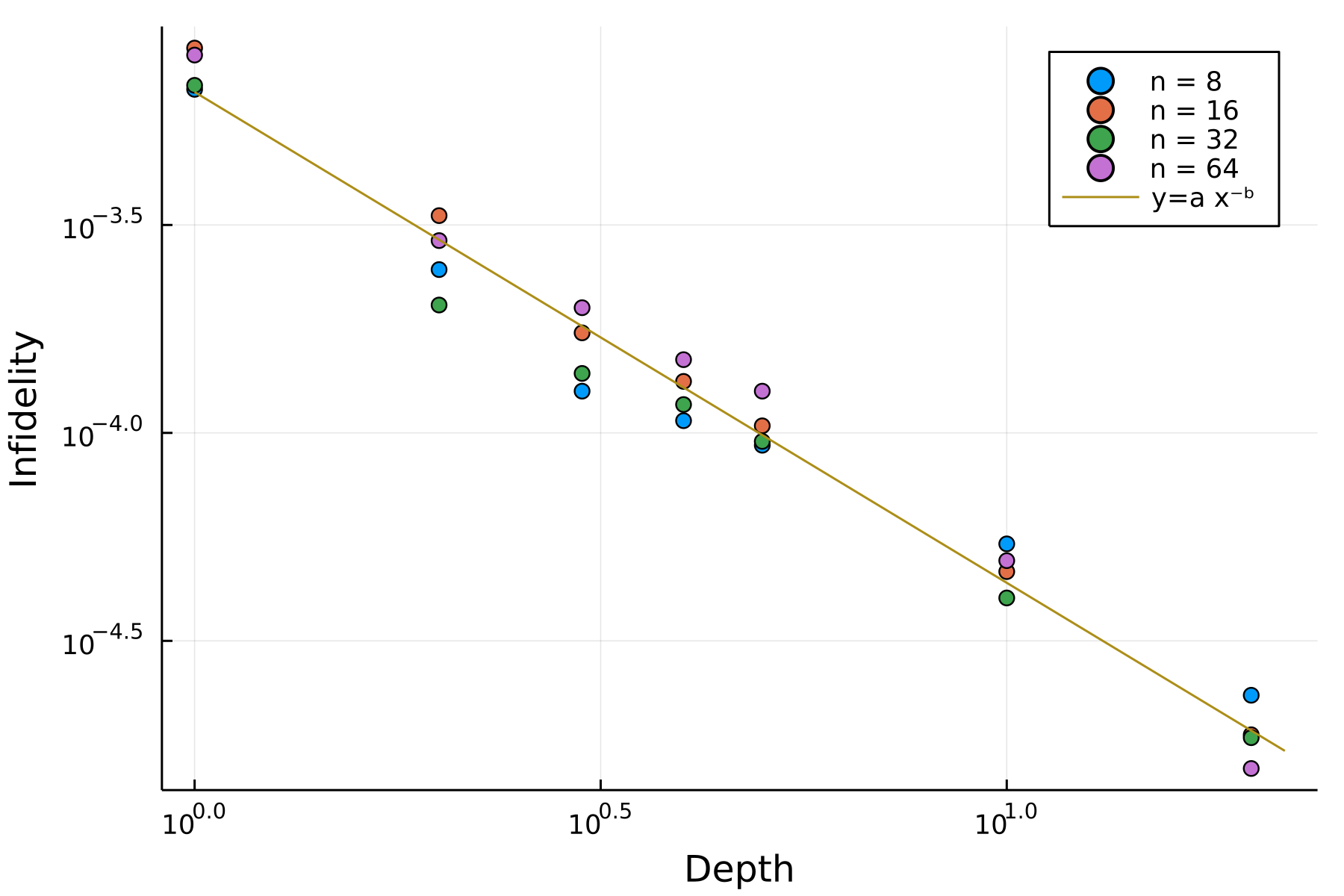}
\caption{{\it(left)} The infidelity between the exact Irwin Hall MPS with bond dimension $\chi \approx n(n+1)$, and the truncated MPS as a function of the truncated bond dimension $\chi'$. We see that, although the results are noisy, the decay is approximately exponential at follows the form $y=ae^{-bx}$ with $a= 6.2$ and $b= 4.5$. {\it(right)} The infidelity between the exact Irwin Hall MPS and the quantum state generated by the iterative MPS approximation circuit as a function of depth $D$. We plot the results for $N=12$. In this case the infidelity decays as a power law $y=ax^{-b}$ with $a= 0.0006$ and $b = 1.15$ }
\label{fig:mps_error}
\end{figure}

In the left plot of Supplementary Fig.~\ref{fig:mps_error}, we see that the MPS error decreases exponentially with the truncated bond dimension $\chi'$.  So while the bond-dimension of the exact Irwin Hall MPS state grows with order n, we only need to generate a quantum state which approximates a relatively small bond-dimension MPS during our state preparation procedure. This quantum state is generated with a quantum circuit of depth D, and there is a separate approximation error associated with how well this output of this quantum circuit approximates the higher bond dimension MPS. In the plot on the right we show the scaling of the overall infidelity with D. We see that the error in this case follows a power law of the form $\epsilon^2 \sim .0006 D^{-1.15}$.  In other words, we require $D \sim 1/\epsilon^{1.74}$. Note however that this exponent is not universal, and in particular we expect that other methods for optimizing the circuit approximation to a MPS, such as those in Ref.~\citeonline{zapata_mps} of the main text can reduce this exponent further. 

\subsection*{Estimation of Errors from Biased Estimators}
We write $\ket{\psi}=\sum\sqrt{p_l}\ket{l}$ and $\ket{\Tilde{\psi}}=\sum\sqrt{\Tilde{p}_l}\ket{l}$, where $p_l$ represents probabilities from the ideal normal distribution $\mathcal{N}$ and $\Tilde{p}_l$ represents probabilities obtained from the approximatd MPS state. We denote the underlying distribution from MPS state as $\mathcal{M}$.

With that, we can express the error from a biased estimator as
\begin{equation}
    |\mathbb{E}_\mathcal{M}[g]-\mathbb{E}_\mathcal{N}[g]|\,\leq\, \int |g(x)|\cdot|f_\mathcal{M}(x)-f_\mathcal{N}(x)|dx 
    \,\leq\, \|g\|_\infty\int|f_\mathcal{M}(x)-f_{\mathcal{N}}(x)| dx,
\end{equation}

Since the random variables are discretized, we can write $\int|f_\mathcal{M}(x)-f_{\mathcal{N}}(x)| dx$ as 
\begin{equation}
    \sum_l|p_l-\Tilde{p}_l|=\sum_l|\sqrt{p_l}-\sqrt{\Tilde{p}_l}||\sqrt{p_l}+\sqrt{\Tilde{p}_l}|\leq \left(\sum_l\left|\sqrt{p_l}-\sqrt{\Tilde{p}_l}\right|^2\right)^{1/2}\left(\sum_l\left|\sqrt{p_l}+\sqrt{\Tilde{p}_l}\right|^2\right)^{1/2},
\end{equation}
where we used the Cauchy-Schwarz inequality.

We note that $\left(\sum_l\left|\sqrt{p_l}-\sqrt{\Tilde{p}_l}\right|^2\right)^{1/2}=\|\ket{\psi}-\ket{\Tilde{\psi}}\|\leq \epsilon$, and the second term can be bounded from above as
\begin{equation}
    \left(\sum_l\left|\sqrt{p_l}+\sqrt{\Tilde{p}_l}\right|^2\right)^{1/2}\leq \left(\sum_l 2p_l+2\Tilde{p}_l\right)^{1/2}=2.
\end{equation}

As such, the error from a biased estimator can be expressed as
\begin{equation}
    |\mathbb{E}_\mathcal{M}[g]-\mathbb{E}_\mathcal{N}[g]|\leq 2\|g\|_\infty \|\ket{\psi}-\ket{\Tilde{\psi}}\|\leq 2\|g\|_\infty\epsilon=O(\epsilon).
\end{equation}

\appendix

\end{document}